\documentclass{aastex}
\usepackage{emulateapj5,apjfonts}
\newcommand{\LCDM}{$\Lambda$CDM~}
\newcommand{\cmsGeV}{cm$^2\;\mbox{GeV}^{-1}$~}
\newcommand{\gmcmc}{gm$\;\mbox{cm}^{-3}$~}
\newcommand{\cms}{cm$\;\mbox{s}^{-1}\;$~}
\newcommand{\cmsgm}{cm$^2\;\mbox{gm}^{-1}\;$~}
\newcommand{\Msunpcc}{$M_\sun\;\mbox{pc}^{-3}\;$~}
\newenvironment{inlinefigure}{%
\def\@captype{figure}%
\noindent\begin{minipage}{0.999\linewidth}\begin{center}}
{\end{center}\end{minipage}\smallskip}

\shorttitle{SIDM AND GRAVOTHERMAL CATASTROPHE}
\shortauthors{Balberg, Shapiro \& Inagaki}

\begin{document}

\submitted{ApJ in press, April 2002}
\title{Self-Interacting Dark 
Matter Halos and the Gravothermal Catastrophe}

\author{Shmuel Balberg\altaffilmark{1,2}, 
Stuart L.~Shapiro \altaffilmark{1,3} and Shogo Inagaki \altaffilmark{4}}

\email{shmblbrg@saba.fiz.huji.ac.il, shapiro@astro.physics.uiuc.edu, 
inagaki@kusastro.kyoto-u.ac.jp}

\altaffiltext{1}{Department of Physics, Loomis Laboratory of Physics,
University of Illinois at Urbana--Champaign, 1110 West Green Street,
Urbana, IL 61801--3080}

\altaffiltext{2}{Racah Institute of Physics,
The Hebrew University of Jerusalem, Givat Ram, Jerusalem 91904, Israel}

\altaffiltext{3}{Department of Astronomy and National Center for
Supercomputing Applications, University of Illinois
at Urbana--Champaign, Urbana, IL 61801}

\altaffiltext{4}{Department of Astronomy, Faculty of Science, 
Kyoto University, Sakyo-ku, Kyoto 606-8502, Japan}

\begin{abstract}
We study the evolution of an isolated, spherical halo of self-interacting dark 
matter (SIDM) in the gravothermal fluid formalism. We show that the thermal 
relaxation time, $t_r$, of a SIDM halo with a central density and velocity 
dispersion of a typical dwarf galaxy is significantly shorter than its age. 
We find a self-similar solution for the evolution of a SIDM halo in the limit 
where the mean free path between collisions, $\lambda$, is everywhere longer 
than the gravitational scale height, $H$. Typical halos formed in this long 
mean free path regime relax to a quasistationary gravothermal density profile
characterized by a nearly homogeneous core and a power-law halo where
$\rho \propto r^{-2.19}$. We solve the more general time-dependent problem and 
show that the contracting core evolves to sufficiently high density that 
$\lambda$ inevitably becomes smaller than $H$ in the innermost region. The 
core undergoes secular collapse to a singular state (the ``gravothermal 
catastrophe'') in a time $t_{coll} \approx 290 t_r$, which is longer than the 
Hubble time for a typical dark matter-dominated galaxy core at the present 
epoch. Our model calculations are consistent with previous, more detailed,
N-body simulations for SIDM, providing a simple physical interpretation of 
their results and extending them to higher spatial resolution and longer 
evolution times. At late times, mass loss from the contracting, dense inner 
core to the ambient halo is significantly moderated, so that the final mass of 
the inner core may be appreciable when it becomes relativistic and radially 
unstable to dynamical collapse to a black hole.

\end{abstract}


\section{Introduction}\label{Sect:Intro}

Cold dark matter has become the paradigm for explaining observations of large 
scale structure of the universe. Flat cosmological models composed of cold 
dark matter (CDM) and a cosmological constant (or quintessence) combined with 
a nearly scale-invariant, adiabatic spectrum of density fluctuations produce 
excellent fits to observed structure on large ($\gg 1\;$Mpc) scales 
\citep{Bahcallal99}. This class of models - collectively labeled as \LCDM~ - 
has received substantial additional support from the recent cosmological 
observations of supernovae at high redshift \citep{SNdataH,SNdataL} and the 
fluctuations in the cosmic microwave background \citep{BOOMERANG,DESAI}. The 
nature and properties of the dark matter remain unknown, however, and present 
one of the greatest challenges of current cosmological research.

A recent suggestion by \citet{SIDM} that observations on galactic and 
subgalactic scales offer unique clues regarding the properties of the dark 
matter has attracted much attention. The basic argument is that the 
``standard'' CDM models, where dark matter particles are assumed to interact 
only through gravity, may be in conflict with some observed features on 
smaller ($\lesssim 1\;$Mpc) scales. In particular, CDM models \citep{NFW97,
Mooreal} tend to predict density profiles of dark matter halos which are 
cuspy and have central densities of 1 \Msunpcc and larger, whereas 
observations of dwarf and low surface brightness galaxies seem to favor 
the presence of a relatively flat core, with typical densities of 
$0.02\;$\Msunpcc \citep{Firmanial01}. Other apparent discrepancies from the 
simulations of CDM are that they tend to predict: a) more than ten times as 
many dwarf galaxies as are observed for the Local Group, b) disks which are 
small and have too little angular momentum, c) triaxial clusters instead of 
relatively spherical ones; see \citet{Daveal01} for discussion and references.
The thrust of the suggestion by \citet{SIDM} is that these possible 
discrepancies may be alleviated if dark matter particles have some additional 
type of interaction, besides gravity, that affects structure formation only 
on these smaller scales\footnote{The proposal by \cite{SIDM} focused on a 
scattering interaction, but the general concept has been expanded by various 
authors, including fluid \citep{Peebles00}, repulsive \citep{Goodman00}, fuzzy 
(Hu, Barkana \& Gruzinuv 2000), decaying \citep{Cen01, Bentoal00} or 
annihilating \citep{RitTka00, KKT00} dark matter.}. 

The physical requirement for 
such ``self-interacting dark matter'' (SIDM) is that on the smaller scales, 
the typical density of dark matter particles is large enough so that 
interactions have a nonnegligible effect, while on large scales the 
probability of interactions is low enough so that the favorable properties of 
CDM vis a vis observed structure are maintained. \citet{SIDM} speculated that 
if the dark matter particles scatter off one another through this interaction, 
the entropy of their phase space must increase while their trajectories will 
become isotropic rather than radial, processes which will lead to a dark 
matter core with a shallower density profile. The ansatz is that a dark 
matter particle in a typical galactic halo should undergo several collisions 
per Hubble time to achieve such halo structure. Since the relevant mean free 
path is $1-1000\;$kpc and typical dark matter densities 
(e.g, in the solar neighborhood) are $\lesssim 1\;\mbox{GeV}\;\mbox{cm}^{-3}$, 
the cross section for SIDM is required to lie in the range $\sigma\approx 
0.8-800\times 10^{-24}\;$\cmsGeV (where 
$1\;$\cmsgm$\approx 1.78\times 10^{-24}\;$\cmsGeV). Later, \citet{Daveal01} 
concluded from numerical simulations that the preferred range for the cross 
section is about $0.5-5\;$\cmsgm. Interestingly, this coincides 
with the range of typical low energy hadronic physics, implying that a 
natural candidate for a SIDM particle could be a light, supersymmetric hadron 
\citep{Wandeltal,BFP01}.

Following the suggestion by \citet{SIDM}, several 
numerical studies of SIDM were conducted via N-body simulations, amended to 
include a Monte-Carlo algorithms for particle-particle scattering. 
\citet{Yoshidaal00} and \citet{Daveal01} examined SIDM structure formation in 
a universe with an initial density fluctuation spectrum. \citet{Burkert00} 
and \citet{KW00} studied the evolution on a single, isolated halo, with 
initial conditions based on the \citet{Hernquist90} model, which describes a 
noninteracting CDM halo (this choice is probably inappropriate for SIDM - see 
below, and also in Dav$\grave{\mbox{e}}$ et al.~2001). 
While there is significant disagreement about 
whether various aspects of the results are favorable for the SIDM model, the 
general trend in the N-body simulations is that the centers 
of halos do indeed become flatter with lower densities with respect to 
standard \LCDM models. 

In this work we revisit the issue of the density profile of a SIDM halo 
by studying the evolution of such a halo via a gravothermal fluid
approximation. In the gravothermal approach, the ensemble of gravitating 
particles is approximated as a fluid in quasi-static virial equilibrium. The 
effective temperature is identified with the square of the one dimensional 
velocity dispersion, and thermal heat conduction is employed to reflect 
the manner in which orbital motion and scattering combine to transfer energy in
the system.  This formalism was originally introduced for the study of 
globular star clusters (see, e.g, reviews by Lightman \& Shapiro 1978 and 
Spitzer 1987), and has proven to be very successful in understanding the 
evolution of these systems \citep{LAR70,HAC78,LBE80}. This 
agreement comes about despite the fact that star clusters are only weakly 
collisional and have long collision mean free paths greatly exceeding the size 
of the cluster, and thermalization is achieved by the cumulative effect of 
repeated distant, small-angle, gravitational (Coulomb) encounters.
In fact, the gravothermal fluid description is much better suited to SIDM 
halos where the thermalizing particle interactions are close encounter, 
large-angle (hard sphere) scatterings. It is reassuring, nevertheless, 
that even in the case of weakly collisional systems like star clusters, 
the gravothermal fluid model does reproduce many of the results found 
from more fundamental analyses of the collisional Boltzmann equation, 
with collisions treated in the
Fokker-Planck approximation \citep{HEN61,SH71,Cohn79,Cohn80,MS80}.

An isolated self-gravitating cluster of particles
in virial equilibrium will relax via collisions 
to a state consisting of an extended halo with a shallow temperature gradient, 
surrounding a hotter, central core region. As time advances, 
the core transfers mass and 
energy through the flow of particles and heat to the extended halo. 
The thermal evolution timescale of the dense core is much shorter than that 
of the extended halo, which essentially serves as a static heat sink. As the 
core evolves it shrinks in size and mass, while its density and temperature 
grow. Increase of central temperature induces further heat transfer to 
the halo, leading to a {\it secular} 
instability on a thermal (collisional relaxation) 
timescale. The secular evolution of the core towards infinite density and 
temperature but zero mass is known as ``gravothermal collapse'' or 
the ``gravothermal catastrophe'' (Lynden-Bell \& Wood 1968; 
see reviews in Lightman and Shapiro 1978 and Spitzer 1987). 
A critical modification to the gravothermal scenario is that a {\it dynamical} 
instability occurs when the particle velocities in the core, or, equivalently, 
when the central potential, becomes relativistic. It is well known that fluid 
stars (i.e.~highly collisional gases) in hydrostatic equilibrium are unstable 
to radial collapse on a dynamical timescale when they are sufficiently compact 
and relativistic. \citet{ZelPod65} originally conjectured and  
Shapiro and Teukolsky (1985a,b, 1986) subsequently demonstrated 
that collisionless gases in virial equilibrium also experience a radial 
instability to collapse on dynamical timescales when their cores are 
sufficiently relativistic. In either case, this dynamically instability will 
terminate the epoch of secular gravothermal contraction and will lead to the 
catastrophic collapse of a {\it finite} mass core, which must end as a black 
hole. Simulations of the catastrophic collapse of relativistic collisionless 
clusters have been performed recently, in part to explore the possibility 
of creating the supermassive black holes (SMBHs) inferred to exist in most 
galaxies and quasars \citep{ShapTeu85c}; see \citet{ShapTeu92} for a review.

In the case of globular star clusters and gravitational scattering, multiple, 
distant, small-angle scattering events dominate over the very rare, close 
encounter, large-angle scatterings until the core is very dense 
\citep{LightShap78}. \citet{Goodman83} showed that the last stage of secular 
core collapse, when large-angle scatterings in the core are not negligible, 
does not differ significantly  from the earlier stages. As we demonstrate 
below, for SIDM there is also a transition between two evolutionary stages, 
but in this case the difference is very significant. These two stages are best 
described by distinguishing between two limits:\\ {\it The Long Mean Free Path 
(lmfp) Limit.} If the typical distance a particle travels between collisions 
is much longer than the gravitational scale height at the particle's position, 
it orbits the cluster center many times unperturbed before being scattered. 
For typical parameters, the low-density, extended halo of a thermalized SIDM 
system will reside in the lmfp regime, while the core may or may 
not. If the core is in the lmfp regime, a particle can escape from the core 
following a single encounter with another particle, since momentum transfer in 
these large-angle encounters is appreciable. The entire volume of the core can 
then participate in heat and mass transfer.\\
{\it The Short Mean Free Path (smfp) Limit.} 
If the core is dense enough, its size may be comparable or even exceed the 
mean free path between collisions. When the mean free path is much smaller 
than the gravitational scale height, particle motion in the core is 
constrained by multiple collisions. In particular, heat conduction is 
obviously less efficient, and transfer of energy and mass from the core to the 
halo is limited to a surface effect at the edge of the core. As demonstrated 
below, even if a SIDM halo is formed initially with a core in the lmfp regime, 
gravothermal evolution will cause the central density to increase and 
ultimately drive the core to a smfp state. We also show that in this smfp 
limit, mass loss from the dense core is severely reduced with respect to a 
lmfp core.

In this paper we apply the gravothermal model to study the evolution of an 
isolated, spherical halo of SIDM. Although this is clearly a simplified
description omitting many of the details of structure formation (mergers, 
accretion, angular momentum transfer, baryon content), our approach offers a 
simple tool for assessing the SIDM proposal, and provides
some new predictions regarding the evolution and late-time gravothermal
collapse of such a system. The model also provides physical insight 
for interpreting the results of more detailed N-body simulations, and serves as
a probe for studying details of the core profile not easily resolved by
these simulations and the late-time behavior whose diagnosis is 
beyond their current reach.

In \S~\ref{sect:TIMES} we estimate 
relaxation timescales characterizing  
typical SIDM halos and compare them with those of globular star clusters. 
In \S~\ref{sect:GTevolve} we outline the gravothermal formalism.
The results of our SIDM halo evolution calculation are 
presented in Sections~\ref{sect:SELFSIM} and \ref{sect:HYDRO}. We first show 
in \S~\ref{sect:SELFSIM} that in the lmfp limit there exists a 
self-similar solution for the halo evolution, equivalent to the \cite{LBE80} 
homology solution for star clusters. Then, in \S~\ref{sect:HYDRO} we perform a 
time-dependent gravothermal numerical calculation to examine the full 
evolution, including the later stages where the core moves into the smfp 
regime. We briefly compare our results with previous  N-body simulations 
in \S~\ref{sect:NBODY}, and present our conclusions and a discussion  
in \S~\ref{sect:CONC}.

\section{Timescales}\label{sect:TIMES}

In a globular star cluster, the thermal relaxation time due to multiple, 
small-angle, gravitational (Coulomb) encounters is given by
(see, e.g., Lightman and Shapiro 1978 and Spitzer 1987)
\begin{eqnarray}
t_r(\mbox{GC}) & = & 
\frac{\mathrm{v}_m^3}{15.4 G^2 m^2 n \log(0.4 N)}\; \nonumber \\
& \simeq & 5\times 10^8\mbox{yrs}\;
\left(\frac{N}{5\times 10^4}\right)^{1/2}\left(\frac{m}{M_\sun}\right)^{-1/2}
\left(\frac{R_h}{5\;\mbox{pc}}\right)^{3/2}\;.
\label{eq:t_rGC}
\end{eqnarray}
In equation~(\ref{eq:t_rGC}) $N$ is the number of stars, $m$ is their mass, 
$n$ is their number density, and $\mathrm{v}_m$ is the stellar velocity dispersion, 
related to the total cluster mass $M$ and the half-mass radius, $R_h$ by the
virial relation $\mathrm{v}_m^2 \approx \frac{1}{2}(GM/R_h)$.
In typical virialized star clusters, the ratio of collision mean free path 
to local system scale height $H$ everywhere satisfies the strong inequality
\begin{equation}\label{eq:lamHGC}
\left(\frac{\lambda}{H}\right)_{GC} \approx 
\left(\frac{t_r}{t_{d}}\right)_{GC} 
\approx \frac{N}{26\log (0.4N)} \gg 1\;,
\end{equation}
where $t_{d} \approx H/\mathrm{v}_m$ is the local crossing or dynamical 
timescale. As a result of this inequality, these systems are only weakly 
collisional, satisfying the collisionless Boltzmann (Vlasov) equation to high 
approximation as they evolve quasistatically on relaxation timescales. For 
typical globular cluster parameters, the relaxation time is significantly 
shorter than the cluster age ($\approx 10^{10}$ yrs), hence all clusters are 
well thermalized. Indeed the timescale for gravothermal core collapse to a 
singular state in such systems satisfies $t_{coll}(GC) \approx 330 t_r(GC)$ 
\citep{Cohn80} and is comparable to typical cluster ages, so that many, 
initially dense clusters are likely to have undergone complete gravothermal 
collapse \citep{PL78,GH85,Spitzer87}.

In an SIDM halo where thermalization is due to close, large-angle interactions,
the relaxation time is the mean time between single collisions. For a cross 
section per unit mass $\sigma$, this time is
\begin{eqnarray}
t_r(\mbox{SIDM})& = & \frac{1}{a \rho \mathrm{v} \sigma} \nonumber \\
& \simeq & 1.40\times 10^{9}\mbox{yrs}
 \left[\left(\frac{a}{2.26}\right)
       \left(\frac{\rho_c}{10^{-24}\mbox{gm cm}^{-3}}\right)\right. \\
& &    \left.\left(\frac{\mathrm{v}_c}{10^7\mbox{cm sec}^{-1}}\right)
       \left(\frac{\sigma}{1\mbox{cm}^2\;\mbox{gm}^{-1}}\right)\right]^{-1}\;.
\nonumber
\label{eq:t_rSIDM}
\end{eqnarray}
where $\rho_c$ is the central density and $\mathrm{v}_c$ is the central velocity
dispersion. The constant $a$ is typcially of order unity; specifically, 
$a=\sqrt{16/\pi}\approx 2.26$ for particles interacting elastically like 
billard balls (hard spheres) with a Maxwell-Boltzmann velocity distribution 
(Reif 1965, eqs.~(7.10.3), (12.2.8) and (12.2.12)). 

As discussed below, for a typical SIDM halo, the ratio $\lambda/H$  
is likely to be $>1$ at formation throughout the halo; with
evolution, this ratio does not change much in the halo, but ultimately drops 
well below unity in the central core.  At this point the system consists of
a fluid core surrounded by a weakly collisional extended halo. By construction,
the typical SIDM halo relaxation time is smaller than a Hubble time, enabling 
the system to thermalize. We show below that a typical halo has a gravothermal 
collapse time $t_{coll}(SIDM) \approx 290 t_r(SIDM)$. Accordingly, halos
forming in the present epoch are in no danger of collapsing, but halos formed 
earlier at high and moderate redshift may have already undergone gravothermal 
collapse.

The assumption of hard sphere interactions adopted here provides the simplest 
description of dark matter particle interactions and is the basis of most 
N-body studies to date. However, more complex interactions, characterized by 
cross sections which may be energy dependent and/or non-elastic, are also 
possible and may yield different evolutionary tracks than the ones we will 
derive.

\section{The Gravothermal Model for Evolution of a SIDM Halo}
\label{sect:GTevolve}

The fundamental equations describing a spherical, virialized gravothermal 
fluid are mass conservation, hydrostatic equilibrium, an energy flux equation  
and the first law of thermodynamics. The latter equation is time-dependent and 
drives the quasistaticevolution. Denoting $M(r)$ as the mass enclosed by 
radius $r$, $\rho(r)$ as the density, $\mathrm{v}(r)$ as the one-dimensional 
velocity dispersion and $L(r)$ as the luminosity through a sphere at $r$, 
these equations are (Lynden-Bell \& Eggleton 1980):
\begin{equation}\label{eq:mass}
\frac{\partial M}{\partial r}=4\pi r^2 \rho
\end{equation}
\begin{equation}\label{eq:HYDstat}
\frac{\partial (\rho \mathrm{v}^2)}{\partial r}=-G\frac{M \rho}{r^2}
\end{equation}
\begin{equation}\label{eq:flux_GEN}
\frac{L}{4 \pi r^2}=-\kappa \frac{\partial T}{\partial r}
\end{equation}
\begin{eqnarray}\label{eq:firstlaw}
\frac{\partial L}{\partial r} & = & - 4\pi r^2 \rho 
\left\{\left(\frac{\partial}{\partial t}\right)_{\!M}\frac{3 \mathrm{v}^2}{2}+
      p\left(\frac{\partial}{\partial t}\right)_{\!M}\frac{1}{\rho}\right\}= \\
& & -4\pi r^2 \rho \mathrm{v}^2 \left(\frac{\partial}{\partial t}\right)_{\!M}
\log\left(\frac{\mathrm{v}^3}{\rho}\right)\;. \nonumber
\end{eqnarray}
We can define a pressure, $p=\rho \mathrm{v}^2$, which appears in the equation of 
hydrostatic equilibrium, and a temperature $k_B T = m \mathrm{v}^2$ which appears in 
the flux equation; here $m$ is the particle mass. The last equality in 
equation~(\ref{eq:firstlaw}) introduces an effective entropy,
\begin{equation}\label{eq:entropy}
s=\log\left(\frac{\mathrm{v}^3}{\rho}\right)\;.
\end{equation}
Note that the time derivatives in equation~(\ref{eq:firstlaw}) are Lagrangian.

The detailed form of the flux equation depends on the nature of the heat 
conducton. In standard heat diffusion, where the mean free path between 
collisions is significantly shorter than the system size, the flux equation is 
\begin{equation}\label{eq:flux_smfp}
\frac{L}{4 \pi r^2}=-b \rho \mathrm{v} \lambda 
\frac{\frac{3}{2}\partial \mathrm{v}^2}{\partial r}\;
\end{equation}
where is $b$ is an effective ``impact parameter'' of order unity and $\lambda$ 
is a collisional scale for the mean free path given by
\begin{equation}\label{eq:lambda_SIDM}
\lambda = \frac{1}{\rho \sigma}\;.
\end{equation}
For a gas of hard spheres with a Maxwell-Boltzmann distribution, the mean 
free path is actually $\ell=\lambda/\sqrt{2}$ (Reif 1965, eq.~(12.2.13)). The 
coefficient $b$ can be calculated to good precision from transport theory, 
and has the value of $b\approx 25\pi/(32\sqrt{6})\approx 1.002$ 
(Lifshitz \& Pitaevskii 1979, chapter 1 eq.~(7.6) and problem 3). 
{\it Note added in proof} - a similar formualtion for the fluid limit of SIDM 
was presented by \citet{GneOst01} who studied substructure in 
galaxy clusters.

Equation (\ref{eq:flux_smfp}) does not apply for a dilute gravothermal system 
where the mean free path is significantly larger than 
the gravitational scale height of the system $H$, where
\begin{equation}\label{eq:scaleheight}
H\equiv\left[\frac{\mathrm{v}^2}{4\pi G \rho}\right]^{1/2}\;.
\end{equation}
In this case, particles make several orbits between collisions, varying their 
radial position by about one scale height. If $\lambda\gg H$, the mean time 
between collisions, $t_r$, is then larger than the dynamical time, 
$t_d\equiv H/\mathrm{v}$, and the flux equation in this lmfp limit can be 
approximated as 
\begin{equation}\label{eq:flux_lmfp}
\frac{L}{4 \pi r^2}=-\frac{3}{2}b \rho \frac{H^2}{t_r} 
          \frac{\partial \mathrm{v}^2}{\partial r}\;.
\end{equation}
\citep{LBW68,Spitzer87}.
The two forms of the flux equations can be combined into a single expression 
in order to treat a general gravothermal system, obtaining
\begin{equation}\label{eq:flux_both}
\frac{L}{4 \pi r^2}=-\frac{3}{2} b \rho \mathrm{v}
\left[\left(\frac{1}{\lambda}\right)+
\left(\frac{\mathrm{v} t_r}{H^2}\right)\right]^{-1} 
\frac{\partial \mathrm{v}^2}{\partial r}\;.
\end{equation}

This equation correctly reduces to Equation~(\ref{eq:flux_smfp}) in the smfp 
limit and to Equation~(\ref{eq:flux_lmfp}) in the lmfp limit. Substituting in 
Equation~(\ref{eq:t_rSIDM}) and Equation~(\ref{eq:lambda_SIDM}) yields
\begin{equation}\label{eq:flux_eq}
\frac{L}{4 \pi r^2}=-\frac{3}{2} a b \mathrm{v} \sigma 
\left[a\sigma^2+\frac{4\pi G}{\rho \mathrm{v}^2}\right]^{-1} 
\frac{\partial \mathrm{v}^2}{\partial r}\;.
\end{equation} 

This combined form of the flux equation arises from our need to calculate heat 
diffusion from the dense core to the dilute, extended halo. While the extended 
halo always resides in the lmfp limit, the character of the core may change. 
Consider the ratio  
\begin{equation}\label{eq:Hvslambda}
\frac{\lambda}{H}=\left(1/\rho\sigma\right)\left(\mathrm{v}^2/4\pi G \rho\right)^{-1/2}=
\left(4 \pi G\right)^{1/2}\left(\mathrm{v}^2\rho\right)^{-1/2}\sigma^{-1}\;.
\end{equation}
Assume that initially the core is dilute enough so that immediately following 
virialization it resides in the lmfp limit where the ratio is above unity.
During gravothermal evolution the core parameters $\mathrm{v}_c$ and $\rho_c$ both 
increase, decreasing the ratio $\lambda/H$. Now the core density 
increases much faster with time than the core temperature 
($d\log(\mathrm{v}_c^2(t))/d\log(\rho_c(t))\approx 0.1$, see below). Hence, a 
nonrelativistic core will enter the smfp regime well before gravothermal 
contraction drives it to a relativistic state and dynamical collapse.  
Therefore,  late in the evolution of a SIDM halo, its core will be 
in the smfp regime, while the extended halo will remain in the lmfp regime and
some transition region must exist between the two.

\vspace{0.5cm}
\section{Self-Similar Solution for Evolution in the lmfp Limit}
\label{sect:SELFSIM}

The equations (\ref{eq:mass}, \ref{eq:HYDstat}, \ref{eq:firstlaw} and 
\ref{eq:flux_eq}) describe the evolution of a spherical, isolated halo of 
SIDM. These equations define a time-dependent diffusion problem, which we 
solve numerically in the next section. However, it is useful and 
instructive to examine first a self-similar solution which can be found for 
these equations in the lmfp limit, following the original 
derivation of \citet{LBE80} for globular clusters.

If the entire halo is sufficiently dilute so that $\lambda\gg H$ everywhere, 
then $a\sigma^2$ can be neglected with respect to $4\pi G(\rho \mathrm{v}^2)^{-1}$ in 
the square brackets in equation (\ref{eq:flux_eq}). This lmfp regime admits a 
self-similar solution which can be obtained by separating variables into time 
and space components as follows:
\vspace{-0.3cm}
\begin{equation}\label{eq:separate}
r=r_c(t)r_*\;,\;M=M_c(t)M_*\;,\;\mathrm{v}=\mathrm{v}_c(t)\mathrm{v}_*\;,\;\rho=\rho_c(t)\rho_*\;,\;
L=L_c(t)L_*\;.\;
\end{equation}
The time-dependent functions give the values of the core parameters
at time $t$. The dimensionless spatial profiles denoted by the asterik ($*$) 
are related by a set of ordinary differential equations, found by substituting
Equation~(\ref{eq:separate}) into the four 
equations~(\ref{eq:mass}-\ref{eq:firstlaw}):
\begin{eqnarray}
\frac{dM_*}{dr_*} & = & r_*^2\rho_* \label{eq:MnoD}\\
\frac{d(\rho_* \mathrm{v}_*^2)}{dr_*} & = & 
-\frac{M_* \rho_*}{r_*^2} \label{eq:HydnoD}\\
\frac{d(\mathrm{v}_*^2)}{dr_*} & = & -\frac{L_*}{r_*^2\rho_* \mathrm{v}_*^3} \label{eq:FluxnoD}\\
\frac{dL_*}{dr_*} & = & r_*^2\rho_* \mathrm{v}_*^2\left[c_1-c_2
\frac{d\log(\mathrm{v}_*^3/\rho_*)}{d\log M_*}\right] \label{eq:FirstnoD}\;.
\end{eqnarray}
The constants $c_1$ and $c_2$ are two eigenvalues of the equations. The system
of four equations plus two eigenvalues is uniquely determined
by specifying six boundary conditions. Four of these conditions 
apply to the origin, where  
\begin{equation}\label{eq:inner}
\rho_*(0)=1,\;\mathrm{v}_*(0)=1,\;M_*(0)=0\; \mbox{and}\; L_*(0)=0\;,
\end{equation}
and the other two conditions arise from the requirement that the extended 
halo remains practically static over the evolution of the core, so that it 
may be treated as infinite. The mass and energy of the extended halo are then 
also infinite, eliminating any scale to the problem and ensuring the 
existence of a self-similar solution. The consequence of this condition on 
the extended halo is that its density profile must tend asymptotically toward 
a power law of the form $\rho_*(r_*\rightarrow \infty) \propto r_*^{-\alpha}$ 
\citep{Spitzer87}. Indeed the condition of self-similarity automatically yields
the relation $\alpha=6(c_1-c_2)/(2c_1-c_2)$ \citep{LBE80}. The asymptotic 
outer boundary 
conditions then arise from equations~(\ref{eq:MnoD}, \ref{eq:FluxnoD}) and are
\vspace{-0.3cm}
\begin{equation}\label{eq:outer}
L_*(r_*\rightarrow \infty)=(\alpha-2)\rho_* \mathrm{v}_*^5 r_*\;\;\;,\;\;\; 
\mathrm{v}_*^2(r_*\rightarrow \infty)=\frac{M_*}{r^*(2\alpha-2)}\;.
\end{equation}

The above derivation is almost identical to the derivation of 
the self-similar solution for star clusters by \citet{LBE80}. 
The sole difference is the form of the flux 
equation (\ref{eq:FluxnoD}), since  
for SIDM $t_r\propto (\rho \mathrm{v})^{-1}$, whereas for globular 
clusters $t_r\propto \mathrm{v}^3/\rho$. 

\begin{inlinefigure}
\centerline{\includegraphics[width=0.97\linewidth]{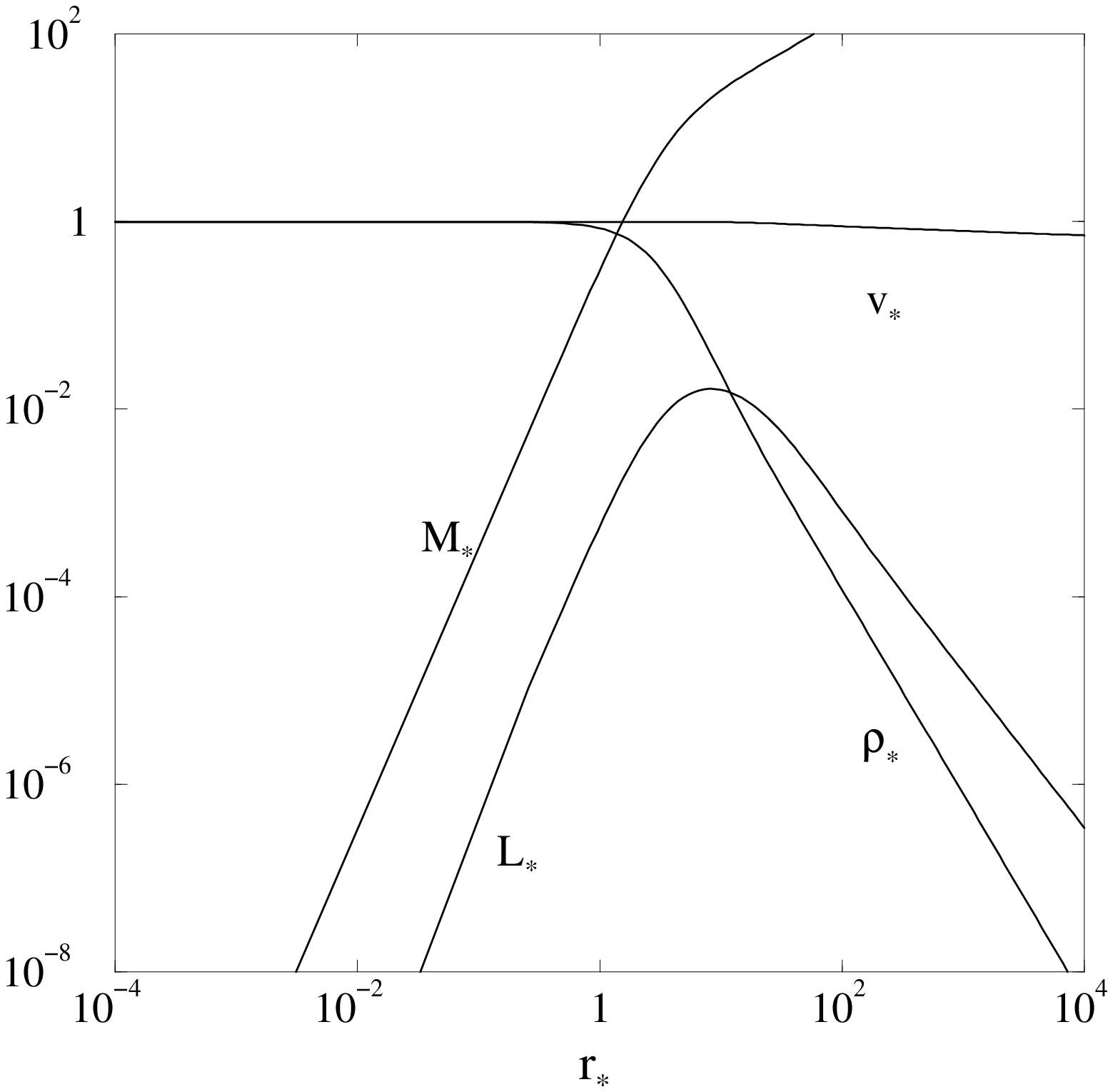}}
\figcaption{The self-similar solution for SIDM 
in the lmfp limit. Mass($M_*$), density ($\rho_*$), 1D velocity dispersion
($\mathrm{v}_*$), and luminosity ($L_*$), are plotted as functions of radius ($r_*$),
all in nondimensional units.
\label{fig:selfsimfig}}
\vspace{0.2cm}
\end{inlinefigure}

We solved the coupled set of ODEs for the nondimensional profiles by 
relaxation on a finite mesh \citep{Eggleton71}, and the solution is shown in 
Figure~\ref{fig:selfsimfig}. A key feature of the solution is the existence of
a core-halo structure: 
the finite core is nearly isothermal and homogeneous, while 
the infinite halo is characterized by declining power-law profiles for
the density, velocity and luminosity. Note that the luminosity 
peaks at the outer edge of the core ($r_*\simeq 1$), reflecting the 
shorter evolutionary (cooling) timescale ($\sim M^2/(L r)$) in this region.

A fundamental feature of the self-similar solution is that the 
eigenvalues $c_1$ and $c_2$ completely define the time-dependent evolution
in terms of $\rho_c(0), \mathrm{v}_c(0)$. We can define an initial core radius 
$r_c(t=0)$ and mass, and $M_c(t=0)$ and follow their time dependence as 
well. Delineating the precise core boundary for the self-similar density 
profile plotted in figure~\ref{fig:selfsimfig} is somewhat arbitrary. We will 
employ the King radius \citep{King66}, where $r_c=3 H(\rho_c,\mathrm{v}_c)$ and then 
define $M_c=M(r_c)$. Typically, the density at the King radius is about $1/3$ 
of the central density.
 
The time-dependent evolution can be parameterized by a single variable
$\zeta$, which is defined through 
\begin{equation}\label{eq:zeta}
\frac{dE_c}{dt}=\zeta\frac{E_c}{M_c}\frac{dM_c}{dt}\;,
\end{equation}
\citep{Spitzer87}
where $E_c$ is the total energy of the core.
Simple dimensional arguments lead to the following relations between 
core mass and other core quantities: 
\begin{equation}\label{eq:dlogdlogM_c}
\frac{d\log r_c(t)}{d\log M_c(t)}=2-\zeta,\;
\frac{d\log \mathrm{v}_c^2(t)}{d\log M_c(t)}=-(1-\zeta),\;
\frac{d\log \rho_c(t)}{d\log M_c(t)}=-(5-3\zeta)\;,
\end{equation}
and $\zeta$ is directly related to the exponent $\alpha$ according to 
\begin{equation}\label{eq:zeta_alpha}
\zeta=\frac{5-2\alpha}{3-\alpha}\;.
\end{equation}
The self-similar solution yields the time dependence of the core quantities in 
terms of the ``collapse time'', $t_{coll}(0)$, the time (from $t=0$) when the 
core mass and radius vanish while the density and velocity dispersion go to 
infinity (relativistic instability is not taken into account). This relation 
is 
\begin{equation}\label{eq:Mcollapse}
\frac{M_c(t)}{M_c(0)} = \left[1-\frac{t}{t_{coll}(0)}\right]^\theta\;,
\end{equation}
where
\begin{equation}\label{eq:tcollapse}
t_{coll}(0) = \theta\frac{1}{\xi_e}t_r(0)\;,
\end{equation}
and
\begin{equation}\label{eq:thetacollapse}
\theta  = \frac{2}{11-7\zeta}\;.
\end{equation}
The mass evaporation rate parameter, $\xi_e$, is defined through the relation 
\begin{equation}\label{eq:xi_e1}
\frac{d M_c(t)}{d t}=-\xi_e\frac{M_c(t)}{t_r(t)},\;
\end{equation}
and is given by  
\begin{equation}\label{eq:xi_e2}
\xi_e=\frac{3}{2}c_2 b\;,
\end{equation}
for the lmfp flux equation (\ref{eq:flux_lmfp}). Similar time-dependent 
expressions for the 
evolution of $r_c$, $\mathrm{v}_c$, and $\rho_c$ follow from 
equations (\ref{eq:dlogdlogM_c}) and (\ref{eq:Mcollapse}).
The sole difference between this solution and the star cluster case arises from
the different functional form of the relaxation time: for Coulomb 
scattering $\theta=2/(7-3\zeta)$. In Table~\ref{tab:parameters} we list 
all the relevant parameters for the self-similar SIDM solution 
and compare them with the star cluster solution \citep{LBE80,Spitzer87}.

\begin{table*}[htb]
\caption{Parameters Defining Self-similar Gravothermal Evolution for SIDM 
and Globular Clusters (GC). 
\label{tab:parameters}}
\begin{center}
\begin{tabular}{l c c c c c c c c c}
\hline\hline
  & $\alpha$ & $\zeta$ & $c_1$ & $c_2$ & $\theta$ & $\xi_e$
  & $\frac{d\log(\mathrm{v}_c^2)}{d\log(M_c)}$
  & $\frac{d\log(\rho_c)}{d\log(M_c)}$ 
  & $\frac{t_{coll}}{t_r}$ \\

\hline

SIDM & 2.190  & 0.7655 & $1.903\!\times\!10^{-3}$ & $8.092\!\times\!10^{-4}$ 
     & 0.3545 & $1.21\!\times\! 10^{-3}$ & -0.2345 & -2.704 & $\sim 291$ \\
GC   & 2.208  & 0.7382 & $2.322\!\times\!10^{-3}$ & $9.704\!\times\!10^{-4}$ 
     & 0.4179 & $1.31\!\times\! 10^{-3}$ & -0.2618 & -2.785 & $\sim 319$ \\
\hline\hline
\vspace{-0.5cm}
\end{tabular}
\end{center}
\end{table*}

Most of the parameters for the SIDM evolution problem have values which are 
very similar to the star cluster case. This is mostly due to the fact that the 
allowed range of the parameters is quite small. Specifically, the exponent 
$\alpha$ must reside between $\alpha=2.0\;(\rightarrow \zeta=1)$, which 
corresponds to an isothermal profile, and $\alpha=2.5\;(\rightarrow \zeta=0)$ 
which corresponds to a constant core energy (see Lynden-Bell \& Eggleton 1980 
for a discussion of this allowed range). In star clusters, transfer of 
particles from the core to the extended halo is a diffusive process, so the 
average change in specific energy should indeed be smaller than in SIDM 
systems, where a particle can be ejected from the core by a single collision. 
Hence, the star cluster values set an upper (lower) limit on the value of 
$\alpha$ ($\zeta$) for the SIDM case, as is borne out by the actual solution. 
An immediate consequence is that the density of dark matter in an isolated, 
extended halo of an SIDM system should maintain the profile  
$\rho(r)\propto r^{-2.19}$. Such a power law is not inconsistent with halo 
profiles previously found with N-body simulations for SIDM. In the same way, 
the predictions of the self-similar gravothermal model for star clusters have
been confirmed by detailed integrations of the Fokker-Planck equation
\citep{HEN61,SH71,Cohn79,MS80}.

The result for $\zeta$ corresponds to the core mass evolving with velocity 
according to the relation $d\log M_c/d\log \mathrm{v}_c^2\approx -4.27$. The core of a 
lmfp SIDM halo must lose more than $99.99\%$ of its mass in order to increase 
its central temperature by one order of magnitude. Assuming that the core is 
initially nonrelativistic, it must increase its central temperature by a 
factor of, say, $10^4-10^6$ to reach the relativistic instability, in which 
case only a tiny fraction of the initial core mass will be available for the 
black hole that arises from the final catastrophic collapse. We show in 
\S~\ref{sect:HYDRO} below how this conclusion is drastically modified once the 
core enters the smfp regime.

One key issue concerning galaxy formation is how the collapse time compares 
with typical ages of galaxies. According to Table 1, a lmfp SIDM system has a 
lifetime of about $290 t_r$ before undergoing secular core collapse. This 
ratio $t_{coll}/t_r$ is somewhat smaller than in the star cluster case, 
because of the greater efficiency of energy transfer in clusters 
as reflected in the value of $b$. The cumulative effect of multiple 
gravitational scatterings in the star cluster case gives $b\approx 0.675$ 
\citep{Spitzer87}, whereas for hard sphere collisions in SIDM systems, we have 
$b = 1.002$. The typical density of dark matter inferred for flat-density 
cores is $\sim 0.02\;$\Msunpcc ($\sim 1.4\times 10^{-24}\;$\gmcmc) 
\citep{Firmanial01}, so for $\mathrm{v}_c\lesssim 10^7\;$\cms the collapse time exceeds 
ten Hubble times (for $\sigma\sim 1\;$\cmsgm). Consequently, SIDM is 
consistent with the existence of relaxed, flat-core density profiles in 
present day halos. Furthermore, the functional forms of 
Equations~(\ref{eq:dlogdlogM_c}) and~(\ref{eq:Mcollapse}) show that throughout 
most of the evolution, the core hardly changes: at 
$t=0.5t_{coll}(0)$ the central density will have increased only by a factor of 
$\sim 2$. Currently existing flat-density cores should exhibit central 
densities similar to those they acquired at virialization.

\vspace{0.2cm}
\section{Time-Dependent Evolution of a SIDM Halo}\label{sect:HYDRO}

We now return to a full time-dependent, numerical solution of the evolution 
equations of \S~\ref{sect:GTevolve}, without assuming self-similarity or 
restricting ourselves to the lmfp regime. The numerical approach is not only  
useful for confirming the self-similar solution, but it is essential for 
studying the evolution of the core in the smfp limit, which it must eventually 
reach. Late in its life (shortly before $t_{coll}$), the core becomes 
sufficiently dense and hot that $\lambda/H$ falls below unity and the 
evolution deviates from the self-similar solution. This final stage will be 
dominated by the thermal evolution of a genuine fluid core, which cools, 
contracts and ultimately becomes dynamically unstable to relativistic collapse 
to a black hole.

It is useful to define a new set of dimensionless 
variables using fiducial mass and length scales, $M_0$ and $R_0$. 
The interaction cross section can then be written as
\begin{equation}\label{eq:sigma_0}
\sigma\equiv\hat{\sigma}{\sigma_0},\;\;
\mbox{with}\;\; \sigma_0\equiv\frac{4\pi R_0^2}{M_0}\;.
\end{equation}
The natural scales for the other variables are 
\begin{equation}\label{eq:charscal} 
\rho_0=\frac{M_0}{4\pi R_0^3},\; \mathrm{v}_0=\left(\frac{GM_0}{R_0}\right)^{1/2},\;
L_0=\frac{G M_0^2}{R_0 t_0}.
\end{equation}
We define the characteristic time scale, $t_0$ as
\begin{equation}\label{eq:t_0}
t_0\equiv\frac{R_0}{a b \mathrm{v}_0 \hat{\sigma}}=\frac{t_{r,0}}{b \hat{\sigma}}\;,
\end{equation}
where $t_{r,0}$ is the relaxation time for $\rho_0,\; \mathrm{v}_0$ and $\sigma_0$. 
This choice allows to us to recast the evolution equation in a  
dimensionless form:
\begin{eqnarray}
\frac{\partial\tilde{M}}{\partial\tilde{r}} & = & \tilde{r}^2 \tilde{\rho}, 
\label{eq:evolmass}\\
\frac{\partial(\tilde{\rho}\tilde{\mathrm{v}}^2)}{\partial\tilde{r}} & = & 
-\frac{\tilde{M}\tilde{\rho}}{\tilde{r}^2},\label{eq:evolhyd}\\
\frac{\partial (\tilde{\mathrm{v}^3}/\tilde{\rho})}{\partial \tilde{t}} & = & 
\frac{3}{2}\frac{1}{\tilde{\rho}}\frac{\partial}{\partial \tilde{r}}
\left[a\hat{\sigma}^2+\frac{1}{\tilde{\rho}\tilde{\mathrm{v}}^2}\right]^{-1}
\frac{\partial \tilde{v}^2}{\partial \tilde{r}} \label{eq:evoldiff}\;.
\end{eqnarray}
In deriving equation~(\ref{eq:evoldiff}) we combined the flux 
equation~(\ref{eq:flux_eq}) and the first law of 
thermodynamics~(\ref{eq:firstlaw}) to eliminate the luminosity and arrive at a 
diffusion equation. The tilde quantities represent the dimensionless 
variables, i.e., $\tilde{x}\equiv x/x_0$. Note that unlike the self-similar 
solution, these generalized equations are not scale-free due to the presence 
of the $a\hat{\sigma}^2$ term in equation~(\ref{eq:evoldiff}).  

By hypothesis, we assume that a SIDM halo is initially dilute enough to be in 
the lmfp limit, so we can use the the self-similar profile
as the initial condition. The natural choices for scaling are then 
$\mathrm{v}_0=\mathrm{v}_c(t=0)$ and $\rho_0=\rho_c(t=0)$, so that 
$\tilde{\mathrm{v}}_c(t=0)=1$,and $\tilde{\rho}_c(t=0)=1$. Hereafter we drop 
all tildes but report results in terms of these dimensionless variables.

We solved this time-dependent diffusion problem numerically by taking 
alternative diffusion and hydrostatic-equilibrium steps. Time dependence is 
followed through equation~(\ref{eq:evoldiff}), and after each successful time 
step the profile is adjusted to maintain hydrostatic equilibrium. Since the 
dynamical time scale of the system is much shorter than the thermal time, 
hydrostatic equilibrium is solved by keeping the entropy of each Lagrangian 
zone fixed, while its position and other thermodynamic properties are 
modified. The effect of the SIDM collisions is examined by varying the value 
of the combination $a\hat{\sigma}^2$. In principle such a (Newtonian) 
calculation can be carried out all the way to collapse, but in practice 
maintaining numerical accuracy  requires constantly refining spatial and 
temporal resolution, which ultimately limits the dynamic range of the 
calculation. Reasonable accuracy can be maintained until to 
$\mathrm{v}_c^2/\mathrm{v}_c^2(t=0)\approx$ a few tens, at which point theb integration time 
step become excessively small. We are able to asses the approximate state of 
the core at the onset of the relativistic instability by extrapolating the 
results of our late integrations to the relativistic epoch.

\subsection{Confirmation of the Self-Similar Solution}
\label{subsect:lmfpcalc}

Our hydrodynamic integrations confirm the self-similar solution for the lmfp 
limit found in \S~\ref{sect:SELFSIM}. By setting $a\hat{\sigma}^2=0$ in 
equation~(\ref{eq:evoldiff}), the entire halo is forced to reside in this 
limit. We show in Figure~\ref{fig:rhosnaplng} snapshots of the 
density profile $\rho(r)$ for $a\hat{\sigma}^2=0$. The initial profile is 
marked by the thicker line, and in the other profiles larger central densities 
correspond to later times. The self-similar structure is clearly maintained 
throughout the evolution, as the core contracts in size and  mass, while the 
extended halo remains largely unchanged. Numerically we find that the density 
in the extended halo falls as $\rho\propto r^{-2.1896}$, corresponding to a 
value of $\zeta\simeq 0.766$. The quantitative fit to the self-similar 
solution is very good, as is the fit for the function $\rho_c(t)$ and other 
quantities (see below).

\begin{inlinefigure}
\centerline{\includegraphics[width=0.97\linewidth]{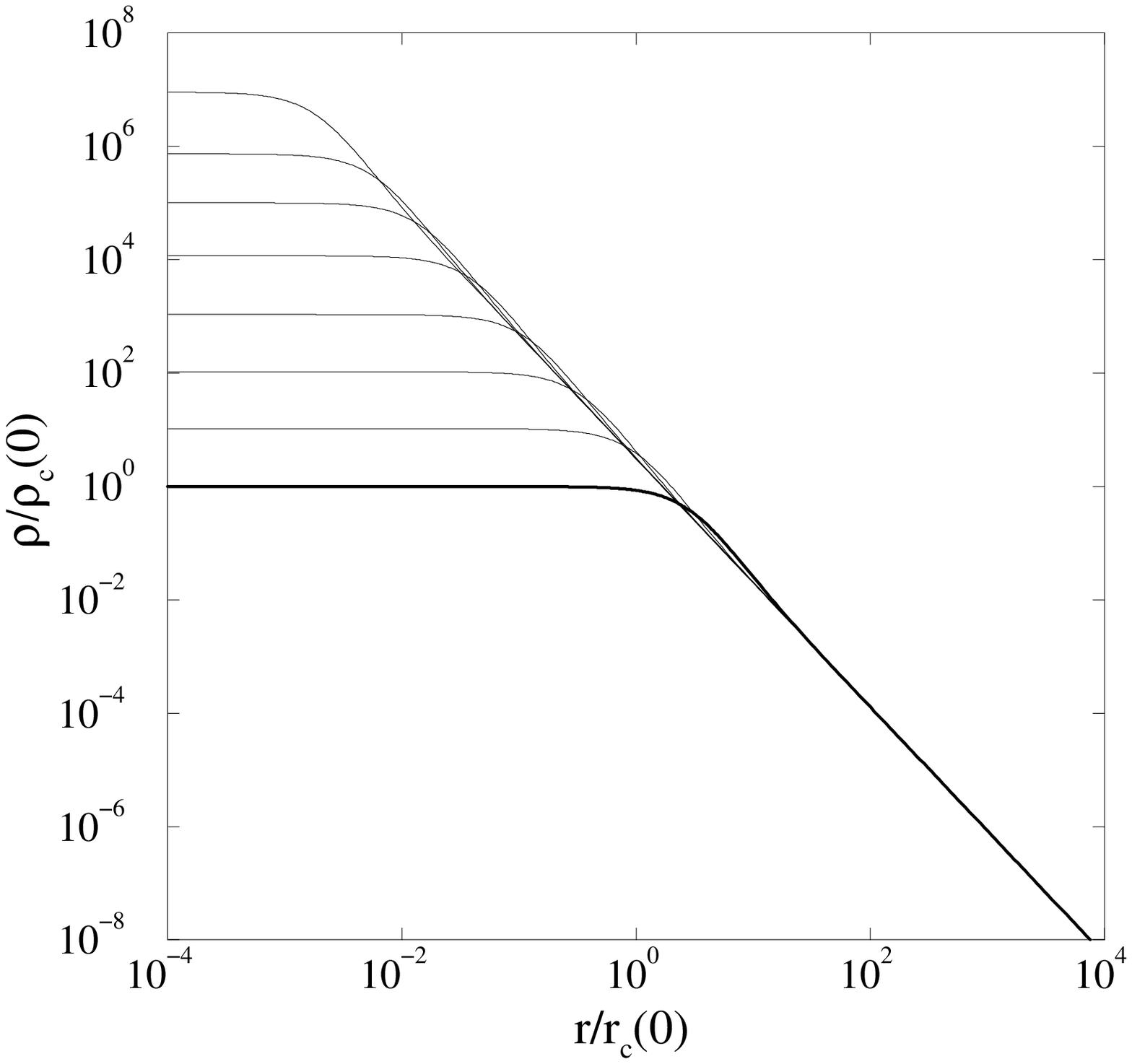}}
\figcaption{Snapshots of the density profile of a lmfp halo of SIDM at 
selected evolutionary times. The thick line denotes the profile at $t=0$; 
larger central densities correspond to later times. 
Profiles are drawn at $t/t_0 = 0.0,\;264.0,\;286.5,\;288.52,\;288.70,
\;288.7178$ and 288.7180.
\label{fig:rhosnaplng}}
\end{inlinefigure}

\subsection{Evolution to a Short Mean Free Path (smfp) Core}
\label{subsect:smfpcalc}

We now consider our results for $a\hat{\sigma}^2>0$, whereby the core 
eventually evolves into the smfp regime. Figure~\ref{fig:rhosnap0.01} 
shows snapshots of the density profile for a calculation with 
$a\hat{\sigma}^2=0.01$; the profiles are labeled 
according to the value of $\lambda/H$ (in units of $a^{-1/2}$) at the center.
At early times, when $\lambda/H>1$ everywhere, the profiles are 
identical to those in figure~\ref{fig:rhosnaplng}, but as the core density and 
temperature increase, the density profile gradually deviates from the self 
similar solution. Some difference is already observable when 
$\lambda/H \lesssim 0.1$, and when the core has $\lambda/H \lesssim 10^{-3}$ 
the density profile has obviously taken a new form. 
In fact, what was originally the core during the lmfp stage of the 
evolution has fragmented into two components: a dense inner core which 
continues to evolve, and an outer region with a lower, roughly uniform 
density,  which connects to the extended halo. This outer core is also almost 
static while the inner core continues to contract with time.

The existence of the double core structure is even more pronounced when 
examining the evolution as a function of the mass coordinate. In 
Figure~\ref{fig:snaprhoT_M} we show snapshots of the density and temperature 
profiles for $a\hat{\sigma}^2=0.01$ (the times of the snapshots are 
not necessarily identical between the two frames). The inner core develops a 
fairly sharp ``edge'', which becomes more distinct as the evolution 
progresses. The nature of this inner core can be understood from the 
$(\lambda/H) (m)$ and $L (m)$ profiles at the late stages of the evolution. 
These are shown in Figure~\ref{fig:TLHlam_last} (along with $\mathrm{v}^2(m)$) for the 
last profile of Figure~\ref{fig:snaprhoT_M}.
The inner core forms in the region where $H\gg\lambda$, and heat conduction 
is dominated by classical diffusion with a short mean free path. Its edge 
corresponds to a transition region where $H\sim\lambda$ and where 
the luminosity peaks. The inner core resembles a fluid star composed of a 
nonrelativistic, nearly isothermal and homogeneous gas undergoing
thermal cooling, surrounded by a thin, exponential envelope. 
The outer part of the core is the remnant of the original 
core in the lmfp limit. This outer core then connects smoothly to the extended 
halo, and barely evolves with time as the inner core contracts.
\begin{inlinefigure}
\centerline{\includegraphics[width=0.97\linewidth]{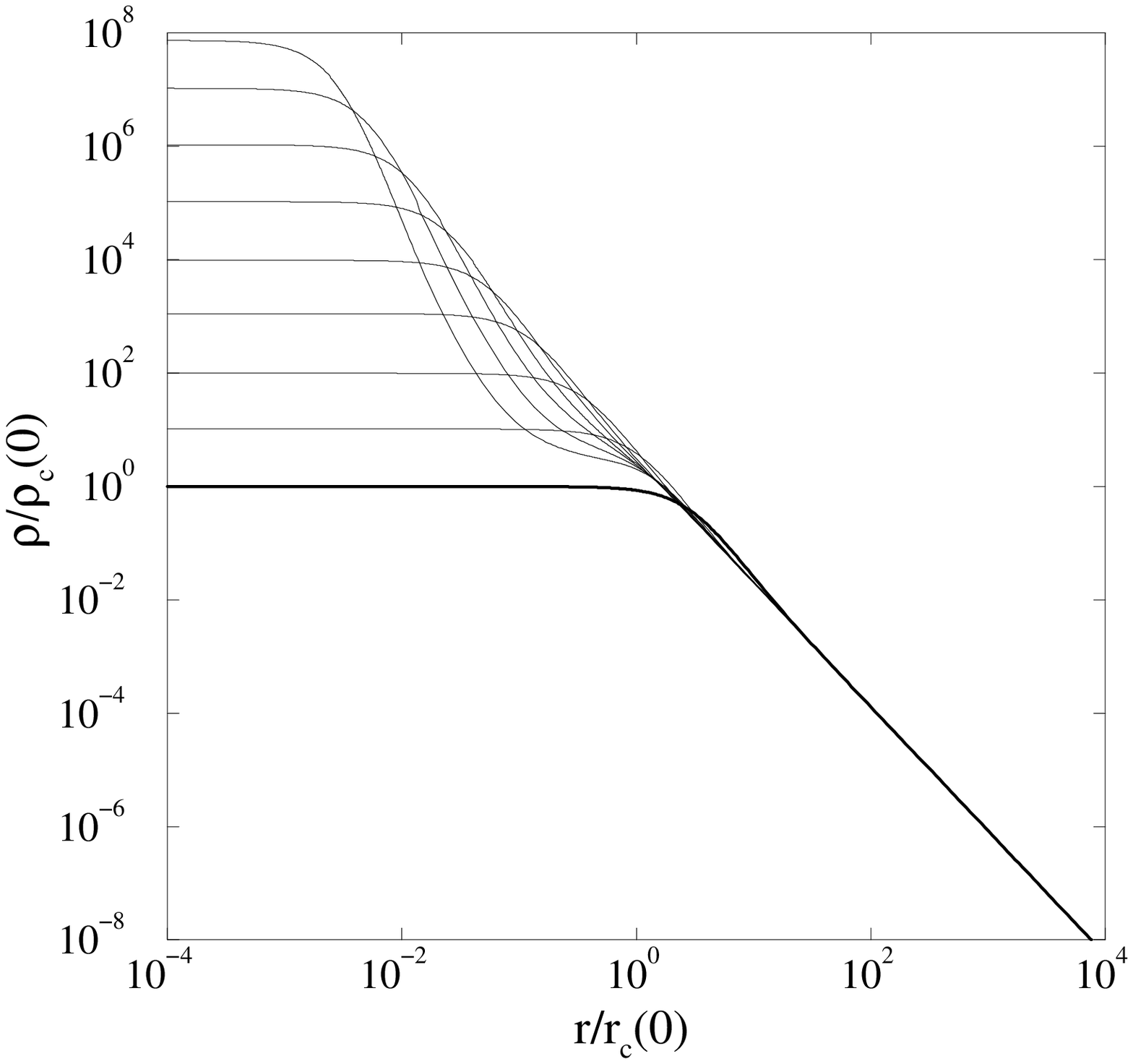}}
\figcaption{Snapshots of the density profile of a SIDM halo with 
$a\hat{\sigma}^2=0.01$ at selected evolutionary times. The thick line 
denotes the profile at $t=0$; larger central densities correspond to later 
times. Profiles are drawn at 
($t/t_0, \lambda/H a^{1/2}$) =
(0.0,10.0), (264.0,2.8), (287.0,$8.2\times 10^{-1}$), 
(289.5,$2.1\times 10^{-1}$), (289.90,$6.0\times 10^{-2}$), 
(290.12,$1.3\times 10^{-2}$), (290.30,$7.0\times 10^{-3}$), 
(290.49,$3.0\times 10^{-1}$), (290.63,$2.1\times 10^{-4}$).
\label{fig:rhosnap0.01}}
\end{inlinefigure}

\vspace{-0.2cm}
An important aspect of the composite core is that the outer region holds most 
of the mass, and therefore acts as a heat sink to the inner region, which 
continues to evolve as its outer layers cool. The inner core loses mass as it 
continues to contract. Gravothermal collapse is slowed down with respect to a 
lmfp regime, where particles orbit outside the core for many
scale heights; in a smfp core, mass and heat loss are now confined to
the surface where $\lambda \approx H$. A slight 
temperature inversion is created in the outer core, which corresponds to a 
region of negative (although very small in magnitude) luminosity. 
The timescale for heat conduction in the outer core is much longer, however, 
than the thermal time of the inner core, so the details of the outer core 
temperature profile are unimportant for the gravothermal evolution.

The value of the cross section, $\hat{\sigma}$, determines the evolutionary 
track of the SIDM halo as its core contracts. The deviation from the 
self-similar solution will begin earlier (at lower central densities and 
temperatures) for larger $\hat{\sigma}$; see equation~(\ref{eq:flux_eq}). A 
larger cross section also reduces the efficiency of heat conduction, since the 
conductivity scales as $\sigma^{-1}$. These effects are demonstrated in 
Figure~\ref{fig:rhos_ts}, which presents the time dependence of the central 
density of a SIDM halo for different values of $a\hat{\sigma}^2$. In all 
cases the initial conditions are the self-similar solution.
In the extreme lmfp limit ($a\hat{\sigma}^2=0$), the collapse time 
appears to be $\sim289 t_r(0)$, in good agreement with the prediction of the 
self-similar solution (assuming $b\simeq 1$). With a finite $a\hat{\sigma}^2$ 
the late-time evolution is slower (requires more relaxation times), as 
expected, and the delay of the late-time evolution is more pronounced for 
larger values of the cross section.  

A finite value of the cross section introduces a scale into the problem which 
precludes a simple, self- similar solution for the evolution of the double 
core and extended halo structure. For example, we show in 
Figure~\ref{fig:rhoc_vc} the relation between central 
density and central temperatures for calculations with different 
$a\hat{\sigma}^2$. The details of the evolution depend on the value of 
$\hat{\sigma}$: there is no universal $\rho_c(\mathrm{v}_c^2)$ relation.
We note in passing that even if the entire system (core and extended halo) 
were in the smfp limit, a self similar solution with a power law density 
profile cannot exist: the thermal time scale in the core would be longer than 
that in the halo, causing the system to be unstable \citep{HAC78}.

\begin{inlinefigure}
\centerline{\includegraphics[width=0.97\linewidth]{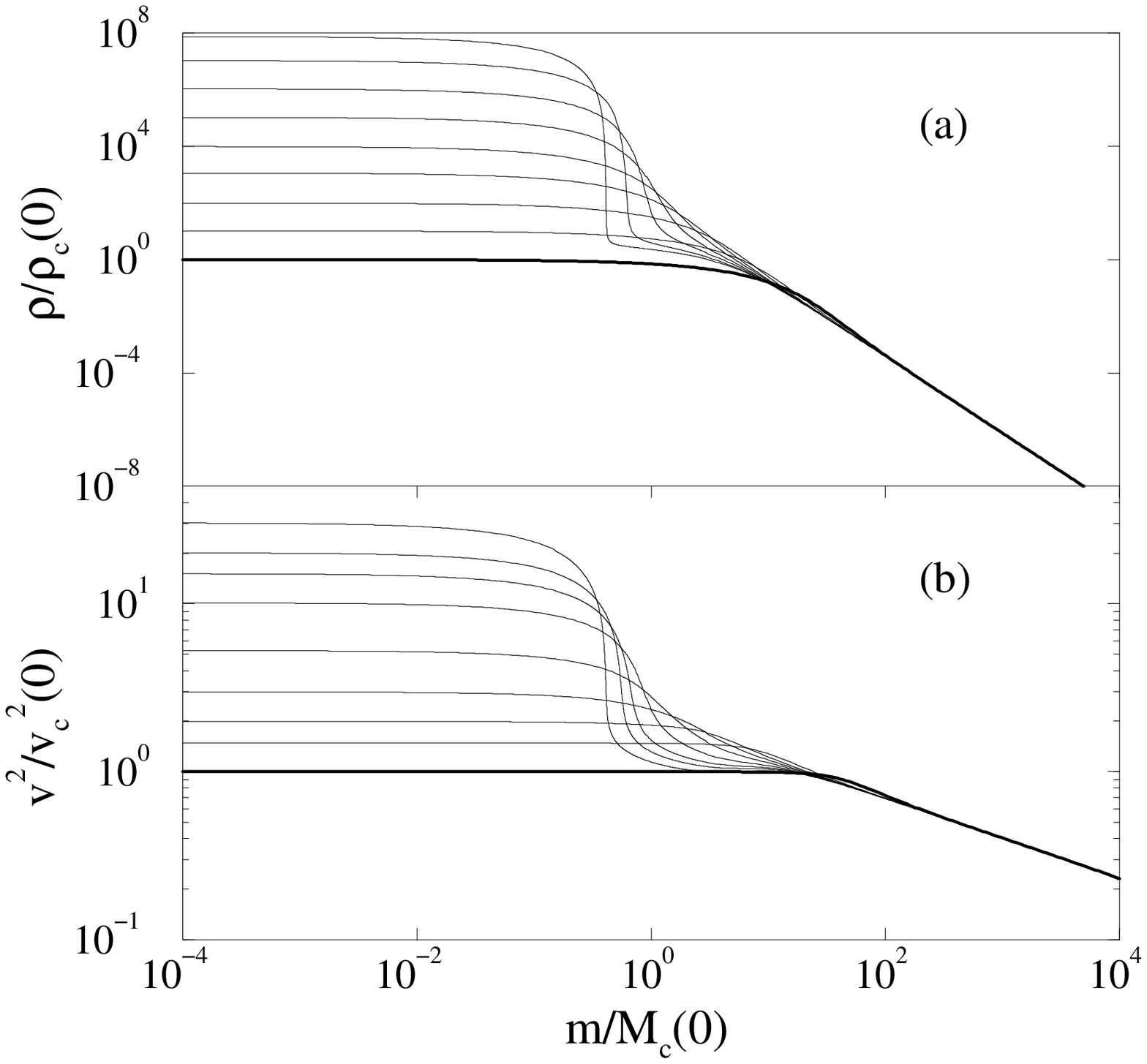}}
\figcaption{Snapshots of the (a) density and (b) temperature profiles of a 
SIDM halo with $a\hat{\sigma}^2=0.01$ as functions of the mass  
at selected evolutionary times. The thick line corresponds to the initial 
conditions. Note the formation of an inner core with 
a relatively sharp surface.
\label{fig:snaprhoT_M}}
\end{inlinefigure}

\begin{inlinefigure}
\centerline{\includegraphics[width=0.97\linewidth]{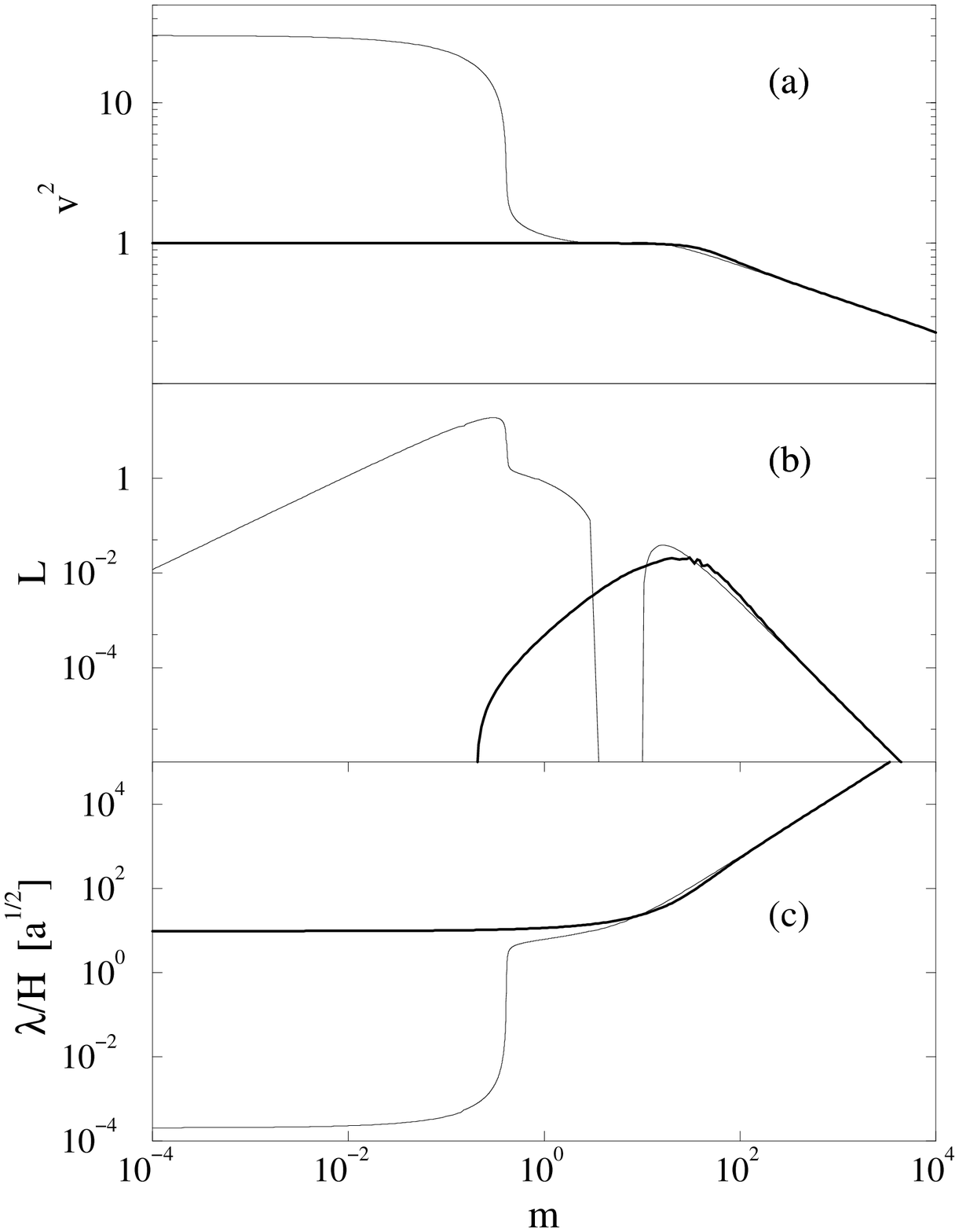}}
\figcaption{The (a) temperature, (b) luminosity and (c) relative mean free 
path $\lambda/H$ profiles of a SIDM halo with $a\hat{\sigma}^2=0.01$ at $t=0$ 
(thick line) and at the termination of the calculation, 
$t/t_r(0)\simeq 290.713$ (thin line).
\label{fig:TLHlam_last}}
\end{inlinefigure}

\begin{inlinefigure}
\centerline{\includegraphics[width=0.97\linewidth]{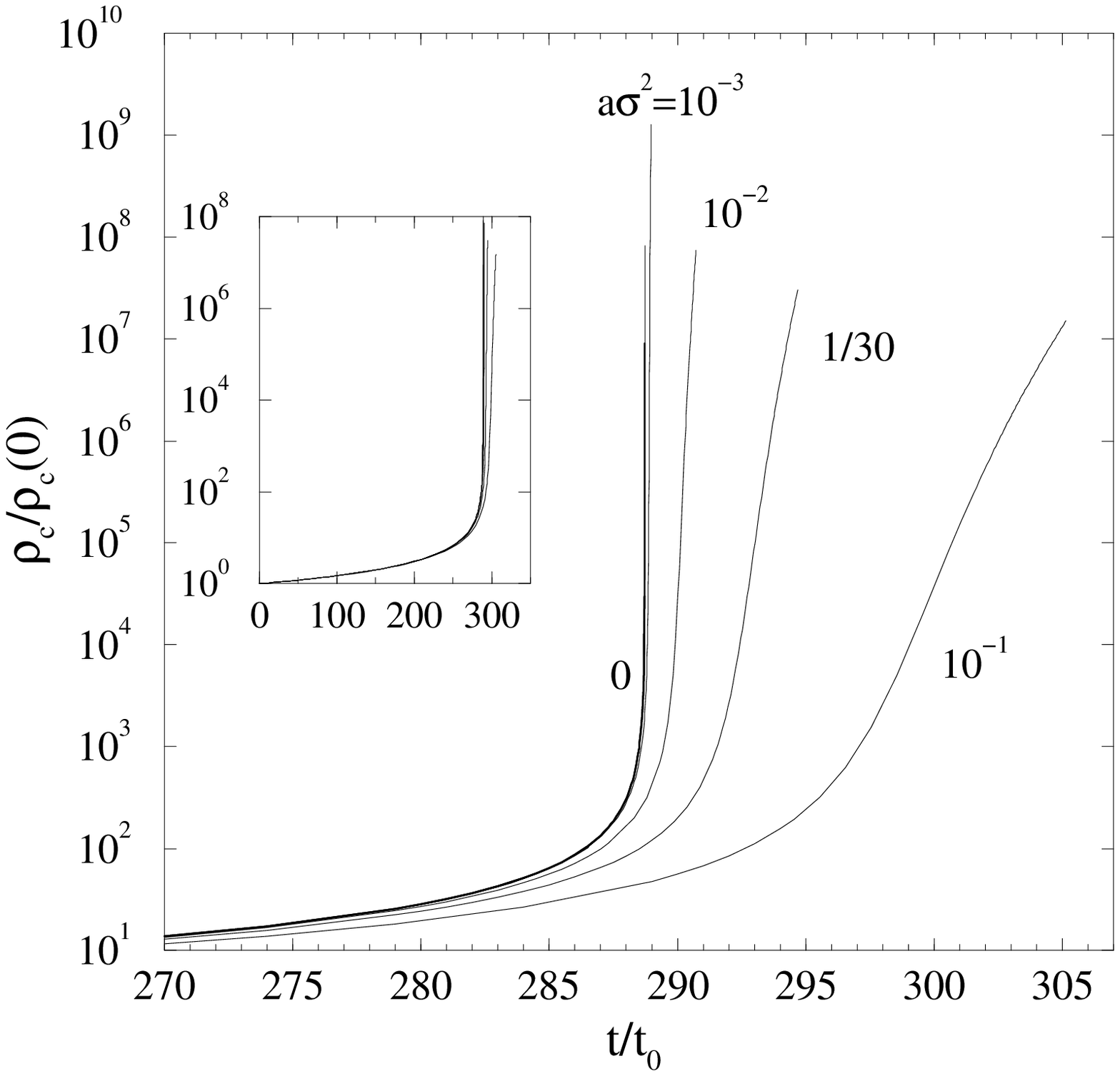}}
\figcaption{Central density as a function of the time for a 
lmfp SIDM halo ($a\hat{\sigma}^2=0$) and for SIDM halos with 
a positive value of $a\hat{\sigma}^2$. The main figure 
shows the late time evolution, when central density increases rapidly; the 
inset shows the entire evolution from $t=0$.
\vspace{0.3cm}
\label{fig:rhos_ts}}
\end{inlinefigure}

\begin{inlinefigure}
\vspace{0.3cm}
\centerline{\includegraphics[width=0.97\linewidth]{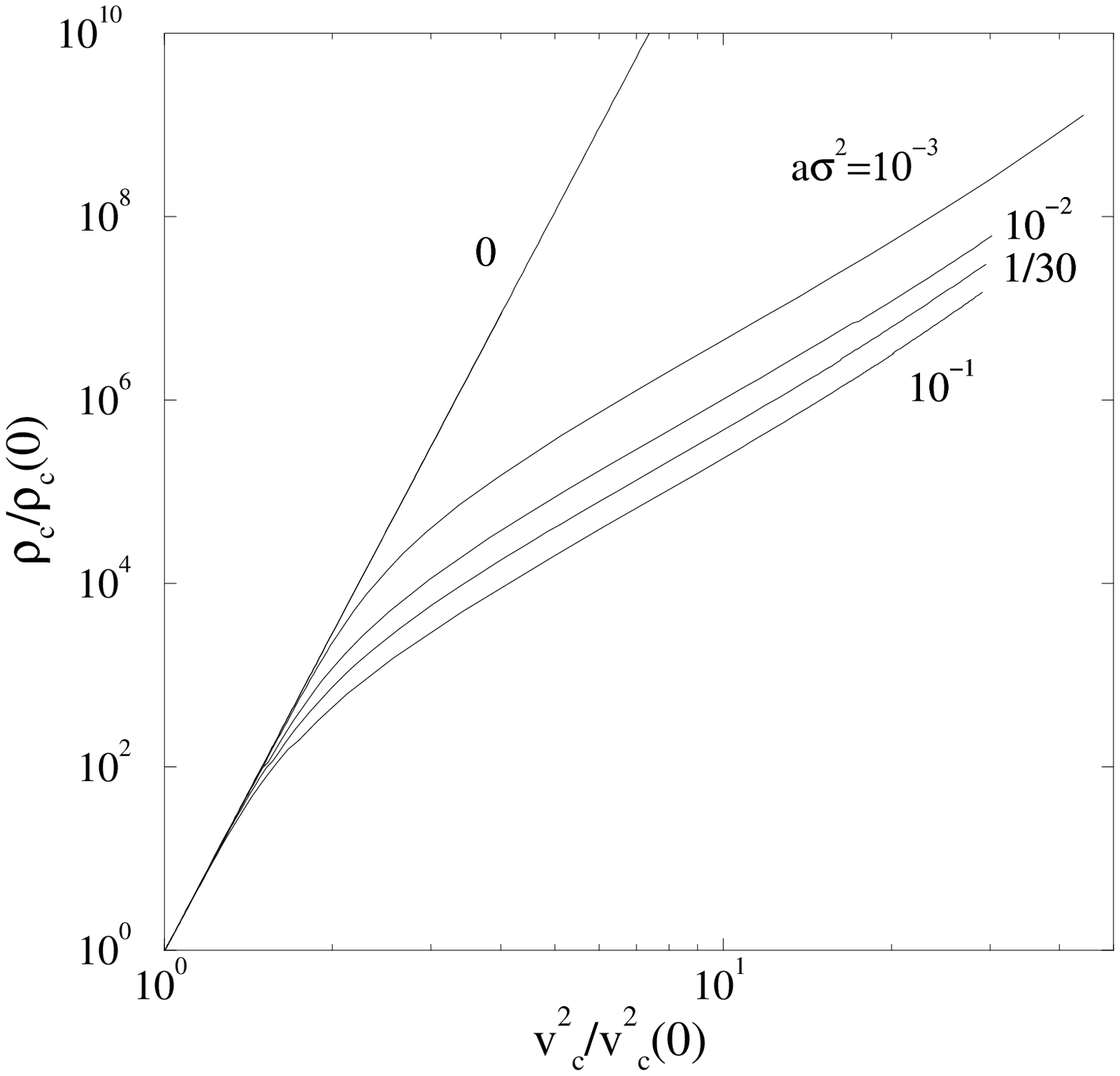}}
\figcaption{Central density vs.~central temperature for a SIDM halo. 
Curves are labled with the value of $a\hat{\sigma}^2$ used in the simulation 
The lmfp self-similar solution is plotted for  $a\hat{\sigma}^2=0$.
\label{fig:rhoc_vc}}
\end{inlinefigure}

The most significant effect caused by the the transition of the inner core 
into a strongly collisional fluid is a critical reduction of the rate of mass 
loss from the core. We demonstrate this in Figure~\ref{fig:Mc_Tc}, which 
shows the core mass as a function of the central temperature 
for the different cases. In the lmfp phase, the core mass 
follows a power law with $d\log M_c/d\log(\mathrm{v}_c^2)\simeq-4.27$, in 
agreement with a value of $\zeta=0.766$ (see eq.~(\ref{eq:dlogdlogM_c})).
When $a\hat{\sigma}^2>0$, the  mass loss rate is appreciably reduced 
for the smfp inner core: instead of a steep power law, the core mass nearly 
saturates while the central temperature increases. As a consequence,
the core mass at the onset of relativistic instability should be much larger 
than would be expected from the self-similar solution in the lmfp limit. 
While strict self-similarity is broken once the inner core enters the smfp
regime, there does seem to exist some power-law correspondance for the 
evolutionary tracks of the central values of some of the physical parameters.
Empirically, we find that the function $M_c/\hat{\sigma}^{0.6}$ 
(in units of $a^{-0.3}$) tends to converge to a single track as a function of 
central temperature; this convergence is also shown in figure~\ref{fig:Mc_Tc}.

\vspace{0.3cm}
\begin{inlinefigure}
\centerline{\includegraphics[width=0.97\linewidth]{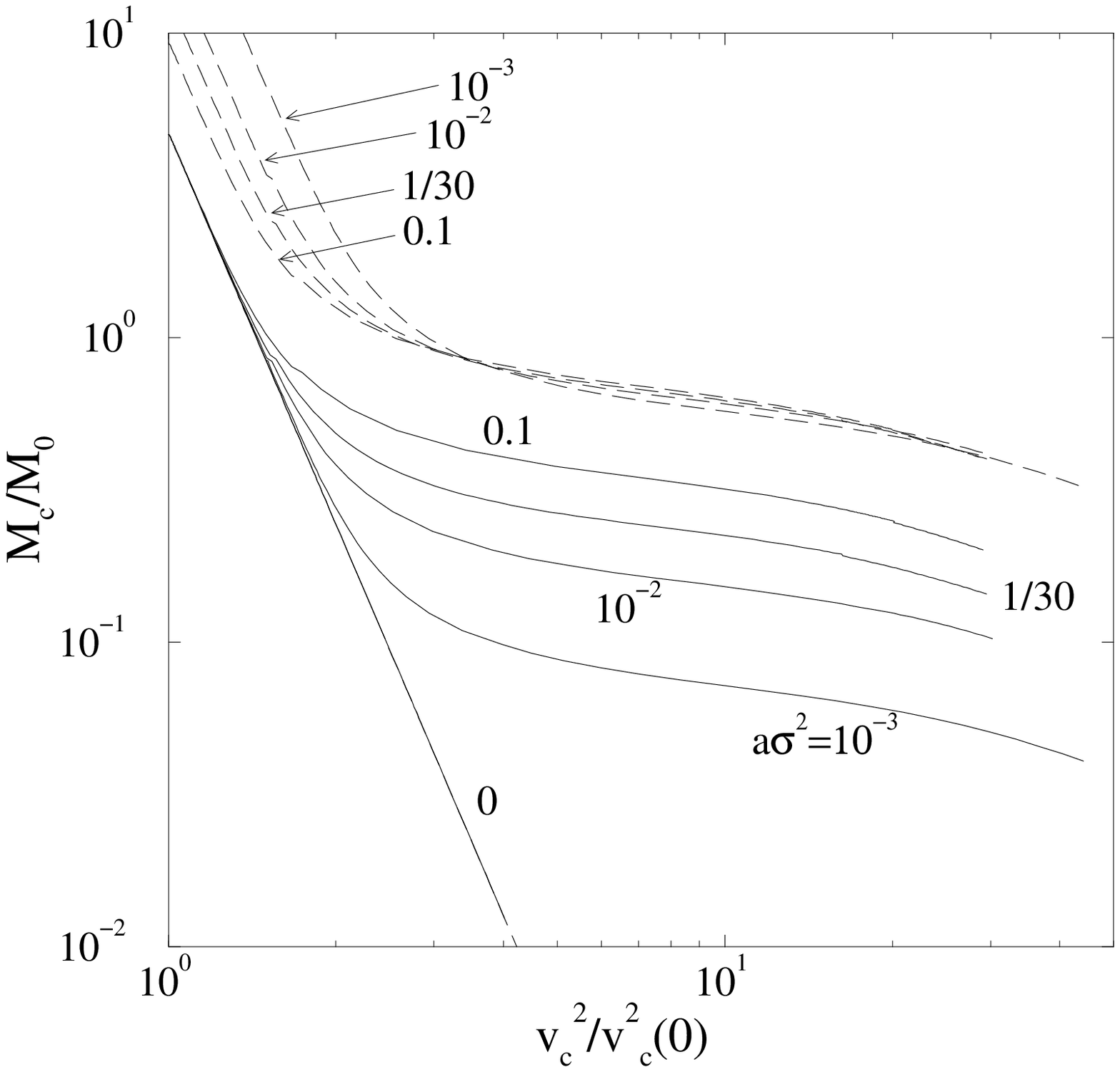}}
\figcaption{The core mass vs.~the central temperature for a SIDM halo 
(solid lines). Curves are labled with the value of $a\hat{\sigma}^2$ used in 
the simulation. The lmfp self-similar solution is plotted for  
$a\hat{\sigma}^2=0$. Also shown are curves normalized as
$M_c/(a\hat{\sigma}^2)^{0.3}$ for the cases with $a\hat{\sigma}^2>0$.
(dashed lines).
\label{fig:Mc_Tc}}
\end{inlinefigure}

\subsection{Late-Time Evolution of a Smfp Core}
\label{subsect:late}

The consequences of relativistic instability in the inner core depend on the 
collisional nature of the gas at the onset of instability. For example, when a 
{\it fluid} star is unstable to radial collapse on a dynamical timescale, it 
undergoe catastrophic collapse in which the entire star (core + halo) forms 
a black hole (see, eg. Shapiro \& Teukolsky 1980).
By contrast, when a {\it collisionless} cluster of particles with a core-halo 
structure is unstable to radial collapse, it is essentially 
the core alone and its immediate surroundings that undergoes catastrophic
collapse (Shapiro \& Teukolsky 1986,1992). Particles in the halo outside the 
core constitute the bulk of the matter and continue to orbit the central black
hole, forming a new, dynamically stable, equilibrium system consisting of a 
central black hole and a massive, extended halo of orbiting particles. 
When a SIDM cluster becomes sufficiently relativistic in the inner core 
it too must be subject to  relativistic instability to collapse, as in the two 
extreme opposite cases described above. At the onset of collapse, the inner
core acts as a fluid system ($\lambda/H \ll 1$) while the exterior region 
behaves as a collisionless gas ($\lambda/H \gg 1$), at least on dynamical 
timescales. This bifurcation is evident in Figure~\ref{fig:TLHlam_last}.
The key consequence is that we expect the entire inner core of a SIDM cluster 
to undergo collapse to a black hole, while the region outside the inner core 
relaxes to a dynamically stable equilibrium system of particles that continue 
to orbit the central hole.

Thus the relevant mass to consider for dynamical collapse and black hole 
formation is the inner core, or, more specifically the region in which the 
matter satisfies $\lambda/H \leq 1$ and behaves as a fluid. The matter inside 
will all collapse and the matter outside this region will reach a stable 
equilibrium state outside the central black hole, at least on dynamical 
timescales. Subsequent interactions between particles in this ambient region 
will ultimately fuel the central hole, causing it to grow on relaxation 
timescales \citep{Ostri00}. The same effect drives the secular growth of 
black holes immersed in ambient star clusters, a problem that has been
well studied by detailed Fokker-Planck calculations (Bahcall \& Wolf 1976, 
Frank \& Rees 1976, Lightman and Shapiro 1977, Cohn \& Kulsrud 1978, 
Marchant \& Shapiro 1980; see Shapiro 1985 for a review and references).  
It is the catastrophic collapse of the inner (fluid) 
core of a SIDM cluster which naturally provides the initial seed black hole.

The $M_c$ vs. $\mathrm{v}_c^2$ relation for the late-time inner core evolution 
determines the mass of the core at the time of relativistic instability. 
Here we infer that relation by a combination of physical reasoning and 
extraplolaton from our numerical integrations for earlier epochs into the 
relativistic regime. Because energy and mass transfer from the smfp inner 
core are limited to surface phenomenon, the  relative rates of these processes 
must remain significantly reduced with respect to the lmfp core. In 
equation~(\ref{eq:zeta}), the typical absolute value of $\zeta$ during the 
late time evolution of the inner core must be close to zero, since only a 
small fraction of the inner core energy is transferred to the outer core. If 
$|\zeta|\sim 0$ then we have $d\log \mathrm{v}_c^2/d\log M_c\approx -1$: hence, at late 
times the inner core will lose about one order of magnitude in mass for every 
order of magnitude it gains in central central temperature (note that $\zeta$ 
cannot be identically zero, as this value corresponds to no evolution 
whatsover). 

We can calibrate this prediction for the late-time $M_c$ vs. $\mathrm{v}_c^2$ relation 
by calculating the value of $\zeta$ numerically using 
Equation~(\ref{eq:dlogdlogM_c}). In Figure~\ref{fig:zetaK} we show the 
corresponding estimates of $\zeta=d\log \mathrm{v}_c^2/d\log M_c + 1$, using the 
results shown in Figure~\ref{fig:Mc_Tc}.
The deviation of the time-dependent 
value of $\zeta$ from the constant value of 0.766 found for the lmfp limit is 
clear and reflects the formation of the smfp inner core. During its  
formation, the \\
\begin{inlinefigure}
\vspace{0.1cm}
\centerline{\includegraphics[width=0.97\linewidth]{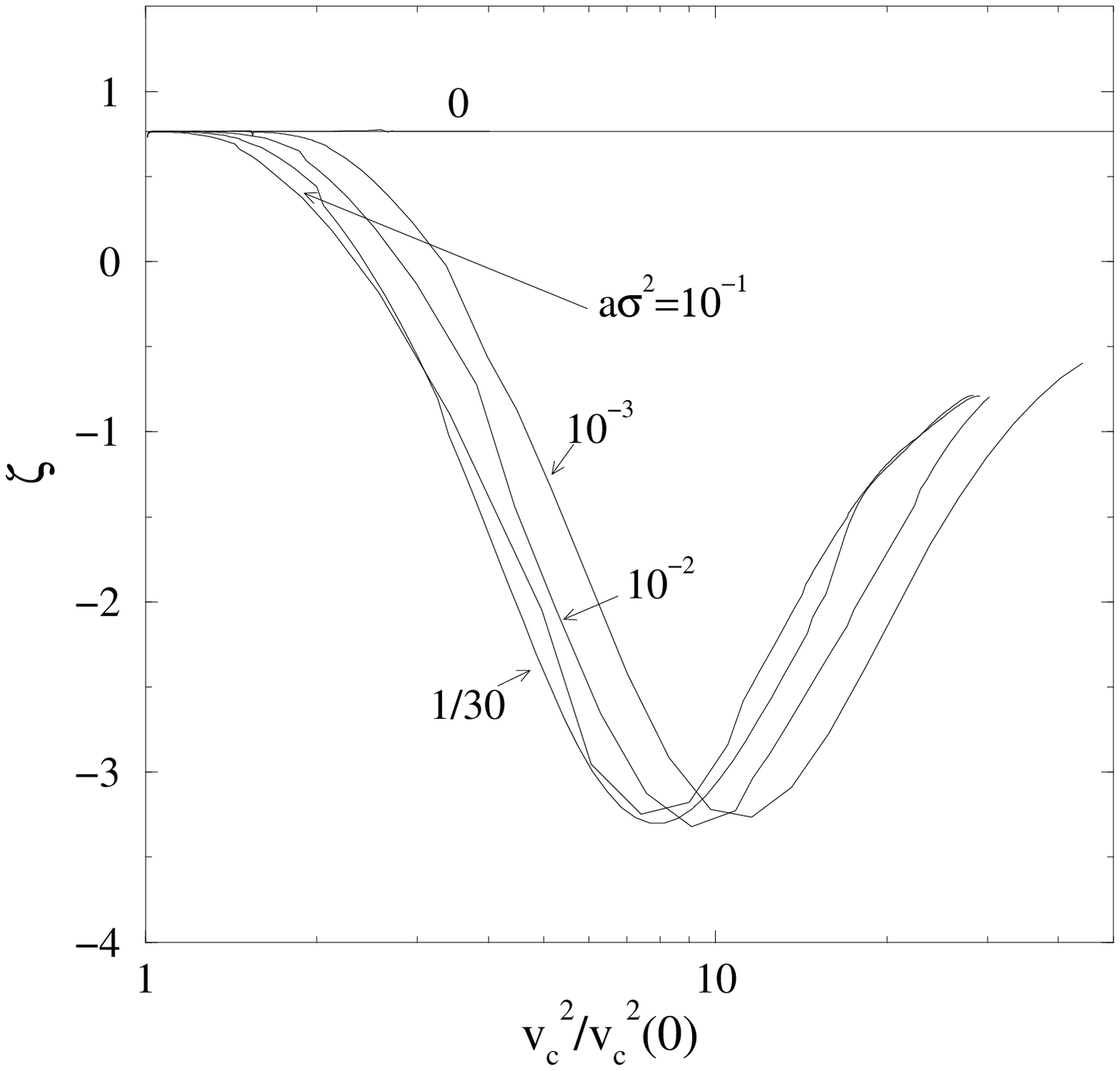}}
\figcaption{The exponent $\zeta$ derived numerically from the relation between 
core mass and the central temperature for a SIDM halo, following 
Figure~\ref{fig:Mc_Tc}.
\label{fig:zetaK}}
\end{inlinefigure}
mass of inner core is almost constant while its density and 
temperature profiles are adjusted, and so $\zeta$ decreases to a minimum 
(negative) value. However, once the inner core is sufficiently dense, 
mass is continuously lost from its surface as outer layers cool and expand to 
join the outer core. The resulting value of $\zeta$ starts to increase again, 
and is still rising when the calculations are terminated. 

We find that due to the sensitivity of the core mass to the details of the 
density profile at the edge of the inner core, it is preferable to estimate 
the asymptotic trend of $\zeta$ by a more robust definition of the core mass. 
In the context of the double core structure where the inner region is
collisional (fluid), the transition point where $\lambda/H=1$ 
provides a more appropriate definition for the ``edge''.
In Figure~\ref{fig:zetaH} we again show the numerical estimate of $\zeta$, 
this time when the core mass corresponds to the radial position where 
$\lambda/H=1$ (which only exisits after a smfp core is formed). Again we see 
that for all cases $\zeta$ is increasing from a transient stage of 
$\zeta\sim -1$, and seems to settle asymptotically in the range 
$-0.2\lesssim \zeta \lesssim -0.1$. We expect that this trend is indeed 
representative of the typical value of $\zeta$ in the case of a smfp core 
at high central temperatures. Our prediction is, therefore, that once 
a smfp inner core is completely formed, the final stages of its evolution 
will be characterized by 
$d\log(\mathrm{v}_c^2)/d\log(M_c)\approx -1.1\sim-1.2$. The 
inner core mass will therefore decrease by only about one order of magnitude 
for every order of magnitude that its central temperature increases.

Finally, our inability to integrate the full system of equations ``forever''
does not prevent us from estimating the the collapse time, $t_{coll}$. In the 
lmfp regime, the collapse time is always $\sim 290$ times the instantaneous 
relaxation time. At the time at which the numerical simulations are halted, 
the smfp inner core relaxation time is typically $10^{10}$ times shorter than 
the initial core relaxation time. Consequently, the total time to collapse 
from $t=0$ is  dominated by the early evolution of the lmfp core and extended 
halo system. The additional time to collapse once the inner core enters the 
smfp regime is considerably shorter. 
\begin{inlinefigure}
\vspace{0.3cm}
\centerline{\includegraphics[width=0.97\linewidth]{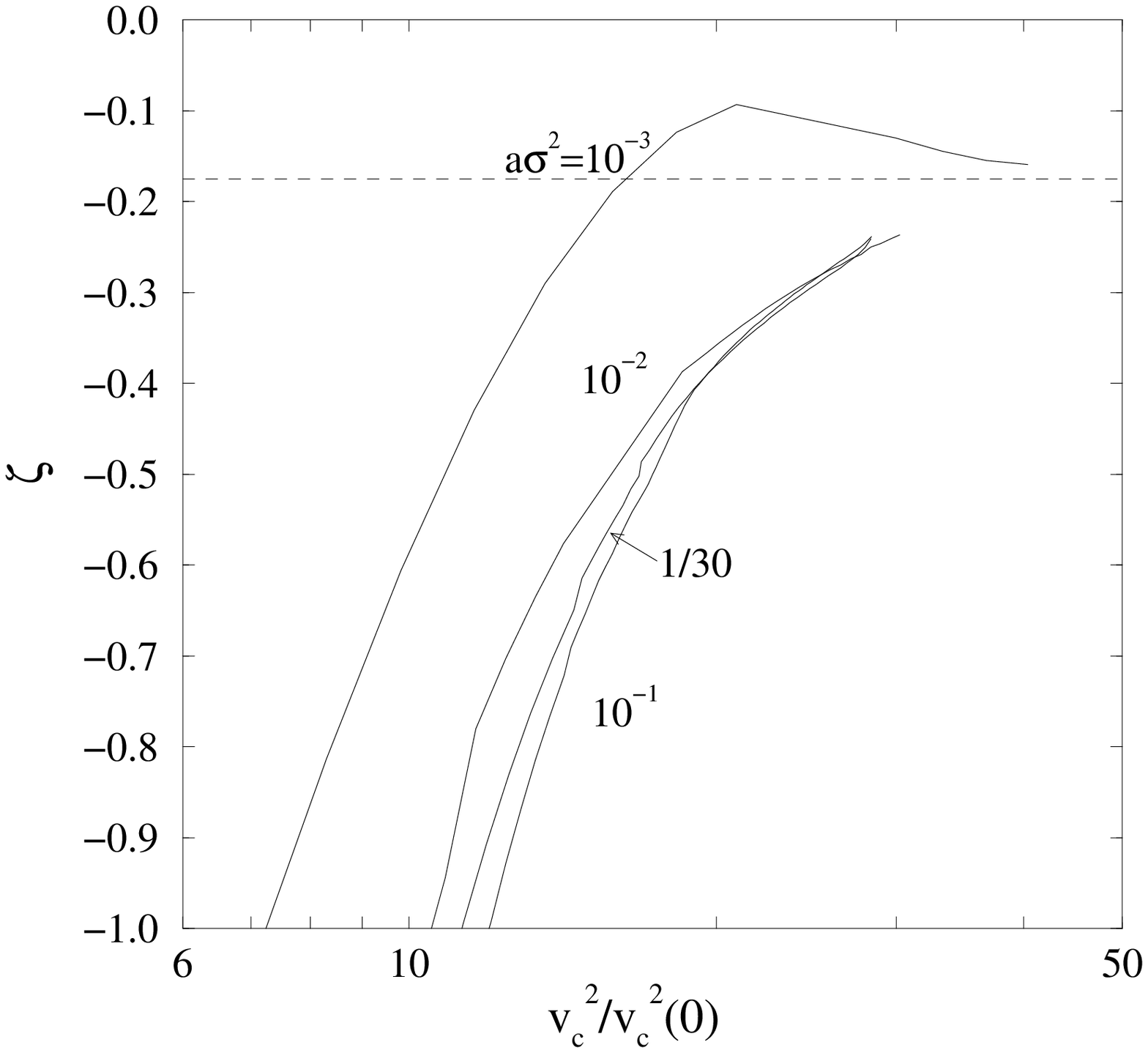}}
\figcaption{The exponent $\zeta$ derived numerically from the relation between 
core mass and the central temperature for a SIDM halo, when the core mass is 
defined by the position where $\lambda/H=1$. The asymptotic value of
$\zeta=-0.175$ at late times is indicated approximately by a horizontal 
dashed line.
\label{fig:zetaH}}
\end{inlinefigure}

\vspace{0.2cm}
\section{Comparison with N-Body Simulations}\label{sect:NBODY}
\vspace{0.2cm}

Our analysis of SIDM evolution via a gravothermal fluid approach is an 
alternative to N-body simulations. In general, mean-field N-body codes have 
the advantage that they avoid having to adopt an approximate, fluid-like set 
of equations to describe SIDM. However, the spatial resolution of N-body codes 
is severly limited by the total number of particles they are able to follow. 
Large numbers of particles, required for high resolution,
can be tracked only for restricted integration times, and this limitation
prohibits N-body simulations from following the evolution of SIDM halos for 
many relaxation timescales.

Previous N-body calculations focussing on the same problem tackled here  
\citep{Burkert00,KW00,Yoshidaal00,Daveal01} also found that introducing 
collisions in the dark matter interactions leads to the formation of a flatter 
core, as opposed to the CDM prediction of a cuspy core (with 
$\rho(r)\propto r^{-\alpha}$, and $\alpha=-1$ in Navarro et al.~1997 or 
$\alpha\simeq -1.5$ in Moore et al.~1999). Previous simulations of 
time-dependent evolution also confirm the general trend of core contraction 
while the extended halo remains roughly static. They also show that
during the evolution the density and temperature of the core both increase 
(see, e.g. in Fig.~1 of Burkert 2000, Figs.~1 and 2 in Kochanek \& White 2000 
and Fig.~3 in Yoshida et al.~ 2000). 

Our study is compared most naturally to the simulations of \citet{Burkert00} 
and \citet{KW00}, who also investigated an isolated halo. \citet{KW00} 
surveyed different values for the SIDM cross section, and discovered that a 
larger cross section induces a greater effect on the evolution - see their 
Fig.~2b, which resembles our Figure~\ref{fig:rhos_ts}. However, both works 
assumed as initial conditions a virialized \citet{Hernquist90} profile with a 
cuspy core, so that initially,  part of the halo resided in the smfp regime: 
the initial conditions of \citet{Burkert00} correspond to a ratio 
$\lambda/H \approx 0.1$ {\it at the outer edge of the core} and those of 
\citet{KW00} correspond to the range of 0.1-3.0. Such a system has a lifetime 
of only a few relaxation times \citep{Quinlan96}, which is the trend found 
by \citet{KW00}, who claimed that such a short lifetime is an argument 
against the validity of SIDM. However, such an initial configuration appears 
to be inconsistent with the SIDM hypothesis, at least for large collision 
cross sections. In this case, a cluster will not virialize with a cuspy halo 
as in the case of CDM, since collisions would modify halo profiles. 
The formation of smooth cores following virialization is certainly indicated 
in the results of \citet{Yoshidaal00} and \cite{Daveal01}, whose calculations 
include halo formation by violent relaxation. Our results may offer a better 
description of the evolution of a SIDM halo: the lifetime of a present day 
halo is significantly {\it longer} than a Hubble time.  

Limited spatial resolution did not allow \citet{Burkert00} and \citet{KW00} to 
follow the evolution of the core to large densities. In particular, the 
simulations they reported were terminated when the central density increased 
by less than a factor of 10 with respect to its value when the flat core 
formed. Thus, they could not observe the late-time evolutionary effects which 
appear at much larger densities, including the formation of the smfp inner 
core or its reduced rate of mass loss as the core continues to contract. We 
hope that N-body simulations or other Boltzmann-solvers with improved spatial 
resolution in the core and extended integration timescales 
might be able to confirm our findings on these issues in the future.
  
We cannot easily compare out results with those of \citet{Yoshidaal00} and 
\cite{Daveal01}, who  examined the significance of a SIDM model on structure 
formation by including a Monte-Carlo module in a cosmological tree code. Their 
calculations are three dimensional and include the effects of accretion and 
mergers, which are not accounted for in our spherical study. \citet{Daveal01} 
claim that no halo ever develops an isothermal core 
since hot material is continuously accreted onto the 
halo, therefore preventing efficient heat transfer from the core to the halo. 
Although their results cannot truly gauge the properties of the core due to 
insufficient resolution, it is clear that significant ongoing accretion 
will modify gravothermal evolution which we found above. However, 
the time-dependent results of \citet{Yoshidaal00} seem to be consistent 
with gravothermal evolution, with the exception of the destabilizing effect 
of major mergers. We plan to extend our method to incorporate an
effective accretion and merger algorithm in the future. 

\vspace{0.3cm}
\section{Conclusions and Discussion}\label{sect:CONC}
\vspace{0.2cm}

The purpose of this work was to construct a simple the gravothermal fluid model
to study the evolution of an isolated, spherical, halo of self-interacting 
dark matter (SIDM). For SIDM every close collision can produce a large-angle 
scattering, a process which can drive thermal relaxation and gravothermal
core collapse. By contrast, CDM halos are collisionless and frozen in time
following their initial virialization. For a typical range of SIDM cross 
sections and central densities of dark matter cores at the present epoch, the 
relaxation time is of order $10^9$ years. Accordingly, a typical halo has had 
sufficient time to thermalize and acquire a gravothermal profile consisting of 
a flat core surrounded by an extended halo. In this sense SIDM is similar to 
globular star clusters, for which the gravothermal model was originally 
developed. However, in globular clusters, relaxation is driven by the 
cumulative effect of many distant, small-angle Coulomb encounters.

In the gravothermal evolution of a SIDM halo, we can  distinguish between a 
lmfp limit, where the gravitational scale height is much smaller 
than the collision mean free path, and the smfp limit, where the situation is 
reversed. Heat conduction and mass transfer between 
the core and the extended halo differ in form and magnitude in the two limits.

A semianalytic self-similar solution exists in the lmfp regime and can be
determined following the approach of \citet{LBE80} for globular star clusters.
The general case requires numerical integration of the full set of 
quasistatic gravothermal equations. We showed that the core of a lmfp halo 
undergoes gravothermal collapse in a time $t_{coll}=290 t_r\;$, where $t_r$ 
is the instantaneous relaxation time. In Newtonian theory the core collapses 
to infinite density and zero mass, but in reality it must collapse slightly 
earlier to a black hole with a finite mass due to the dynamical instability 
which sets in when particle velocities become relativistic. In any case, the 
collapse time of present day halos is significantly longer than the Hubble 
time. Accordingly, the existence of halos with flat cores at the current epoch 
is entirely consistent with the SIDM hypothesis and gravothermal evolution.

In the standard scenario of the gravitational collapse of an initially dilute, 
overdense fluctuation of dark matter, a nascent SIDM halo will virialize by 
violent relaxation on a dynamical (collapse) time scale. The system will 
subsequently thermalize by collisions on a relaxation time scale 
(Eq.~(\ref{eq:t_rSIDM})). For typical parameters, both the core and extended 
halo will be in the lmfp regime following the initial relaxation. The model 
then predicts that the extended halo, if isolated,  will acquire
a density profile with $\rho(r)\propto r^{-2.19}$; this power-law
should remain constant while the core evolves. While a power-law of this 
magnitude is not easily distinguishable from the range considered plausible 
for CDM, $\rho \propto r^{-2}-r^{-3}$, \citep{NFW97,Mooreal98,Kravtsoval98} 
it is consistent with the results of N-body simulations of SIDM. 

While the extended halo remains almost static the core contracts and 
eventually enters the smfp regime where it behaves as a fluid. Subsequent 
evolution drives the core into two components, a dense, smfp inner core and a 
more dilute lmfp outer core. Sharp temperature and density gradients develop 
at the interface between the two components where $\lambda \approx H$.
Gravothermal evolution of the inner core is slowed down (i.e., requires more 
relaxation times) with respect to the evolution of the original lmfp core.

In the lmfp limit we find that $d\log M_c/d\log \mathrm{v}_c^2=-4.27$: the core mass 
must decrease by more than four orders of magnitude for every order of 
magnitude the central temperature increases (very much like the case for 
globular clusters). If at formation the core is initially 
nonrelativistic so that $(\mathrm{v}_c/c)^2\sim 10^{-6}-10^{-4}$, the mass of the 
core at the onset of relativistic instability would clearly be negligible,
were the evolution to persist in the lmfp regime. However, the transition 
into the smfp state drastically changes the evolution, so that 
$d\log M_c/d\log \mathrm{v}_c^2\approx -0.85$. Hence, 
{\it A SIDM halo retains an appreciable inner core mass as it evolves 
towards the relativistic instability}.

Previous N-body simulations of a SIDM halo show that the system develops 
a flat core plus extended halo structure. Studies of an isolated 
halo \citep{Burkert00,KW00} also find that the core evolves by 
contracting in size while increasing its density and temperature, as we 
expect based on our gravothermal model. The gravothermal model allows us to 
follow the evolution of the core to much greater central densities than the 
N-body codes, and  we have able to identify the appearance of a double core 
and reduced mass loss from the contracting inner core. We hope that future 
N-body simulations with improved resolution in the core region 
might be able to confirm this interesting behavior. 

The most significant cosmological implication of our results is that an 
isolated SIDM halo at the present epoch with a central density and velocity 
dispersion similar to those inferred from observations 
($\sim 0.02\;$\Msunpcc and $\lesssim 10^7\;$\cms respectively)
should have a relaxed gravothermal profile, including a flat core and 
an extended halo at the current epoch.  Its relaxation time is less than one 
tenth of Hubble time, so that it had sufficient time to thermalize, while the 
lifetime of its core should exceed ten Hubble times, so the core 
should be far from collapse. Hence, a gravothermal SIDM halo is consistent 
with the observational inference of a flat core in dark matter halos. 
While our approach does not include accretion, mergers or angular momentum,
these processes can be incorporated in more complicated gravothermal models.

One natural question arises: 
did SIDM halos formed at high or moderate redshift earlier in the universe 
have core lifetimes that were shorter than the Hubble time? 
At high red shift the mean density of 
the universe was higher, and so would be the typical densities of halos 
which form, implying shorter relaxation and gravothermal collapse times. 
A scenario whereby the gravothermal catastrophe in a SIDM halo
leads to the formation of a black hole at high red shift is 
intriguing as a general mechanism for producing
supermassive black holes in galaxies and quasars.
We address this possibility in a separate work.

\acknowledgements

We are grateful to P.~P.~Eggleton, E.~Livne, P.~R.~Shapiro, and 
B.~D.~Wandelt for stimulating discussions. This work was supported in part by 
NSF Grant PHY-0090310 and NASA Grants NAG5-8418 and NAG5-10781 at the 
University of Illinois at Urbana-Champaign.

\end{document}